\documentclass[onecolumn]{aastex631}

\usepackage[normalem]{ulem}
\submitjournal{ApJ}
\usepackage{amsmath,amssymb}
\usepackage[caption=false]{subfig}
\usepackage{ulem} 
\shorttitle{M dwarfs' dichotomy}
\shortauthors{Nigro et al.}
\begin{document}
\title{On the Temporal Variability in the Magnetic Dichotomy of Late M Dwarf Stars}

\correspondingauthor{Giuseppina Nigro}
\email{giuseppina.nigro@roma2.infn.it, giusy.nigro@gmail.com}
\author[0000-0001-8044-5701]{Giuseppina Nigro}
\affiliation{Dipartimento di Fisica, Universit\`a degli Studi di Roma Tor Vergata, Via della Ricerca Scientifica 1, Roma, 00133, Italy}
\author[0000-0002-2276-3733]{Francesco Berrilli}
\affiliation{Dipartimento di Fisica, Universit\`a degli Studi di Roma Tor Vergata, Via della Ricerca Scientifica 1, Roma, 00133, Italy}
\author[0000-0002-4896-8841]{Giuseppe Bono}
\affiliation{Dipartimento di Fisica, Universit\`a degli Studi di Roma Tor Vergata, Via della Ricerca Scientifica 1, Roma, 00133, Italy}

\author[0000-0003-2500-5054]{Dario Del Moro}
\affiliation{Dipartimento di Fisica, Universit\`a degli Studi di Roma Tor Vergata, Via della Ricerca Scientifica 1, Roma, 00133, Italy}%

\author[0000-0001-7369-8516]{Luca Giovannelli}
\affiliation{Dipartimento di Fisica, Universit\`a degli Studi di Roma Tor Vergata, Via della Ricerca Scientifica 1, Roma, 00133, Italy}

\author[0000-0002-3948-2268]{Valentina Penza}
\affiliation{Dipartimento di Fisica, Universit\`a degli Studi di Roma Tor Vergata, Via della Ricerca Scientifica 1, Roma, 00133, Italy}

\author[0000-0001-8623-5318]{Raffaele Reda}
\affiliation{Dipartimento di Fisica, Universit\`a degli Studi di Roma Tor Vergata, Via della Ricerca Scientifica 1, Roma, 00133, Italy}


\begin{abstract}
Rapidly rotating late M dwarfs are observed in two different branches of magnetic activity, although they operate in the same stellar parameter range. Current empirical evidence indicates that M dwarfs with spectral types ranging from M3 / M4 to late-type M dwarfs, stellar masses smaller than 0.15 M$_\odot$, and rotational period shorter than four days display either a stable dipolar magnetic field or magnetic structures with significant time variability. The magnetic activity of fully convective M dwarfs is known to be regulated by a mechanism named the ${\alpha}^2$ dynamo. To further constrain the physics of this mechanism, we use a low-dimensional model for thermally driven magnetoconvection producing an ${\alpha}^2$ dynamo, specifically a modified magnetohydrodynamic (MHD) shell model. Although the model neglects density stratification, it captures the essential nonlinear dynamics of an $\alpha^2$ dynamo. Therefore, the results should be interpreted in a qualitative sense, highlighting possible trends rather than providing direct quantitative predictions for fully convective stars. 
The model is validated by comparing the statistical properties of magnetic polarity reversals with paleomagnetic data, since the geodynamo provides the only natural ${\alpha}^2$ dynamo with sufficiently rich reversal statistics. Our findings reveal that increased convective heat transport correlates with more frequent magnetic-polarity reversals, resulting in enhanced magnetic variability. This suggests that the observed magnetic dichotomy in late M dwarfs could be interpreted in terms of differences in global heat transport efficiency. However, additional models and observations of M dwarfs are needed to further constrain this interpretation.
\end{abstract}

%
%



\section{Introduction}
\label{Intro}
Magnetic fields are ubiquitous in the universe and play a fundamental role in the dynamics of many astrophysical objects. They govern the activity of many different types of stars, the environment around planets, and the evolution of very large systems, such as galaxies.  
One of the most intriguing characteristics of astrophysical magnetic fields is their capability to abruptly invert their polarity, producing magnetic reversals - a phenomenon that is still not fully understood. 

Magnetic reversals can exhibit periodicity, as in the approximately 11-year cycle of the Sun \citep{Stix_2004}, or quasiperiodicity, as observed in some late-type stars \citep{Metcalfe_et_al_2013ApJ, Mascareno_et_al_2016A&A, Wargelin_et_al_2017MNRAS, Jeffers_et_al_2022A&A}. In contrast, very low-mass stars \citep{Stauffer_2017AJ} and Earth's magnetic field exhibit more irregular, chaotic behavior. However, in the geodynamo, some evidence suggests the presence of periodic components, possibly related to mantle convection, superimposed on chaotic variability \citep{Driscoll_Olson_2011}. 

The temporal behavior of magnetic fields in astrophysical plasmas, including reversals and the amplification of magnetic field seeds, is governed by a dynamo mechanism. 
This mechanism is driven in the convection zone of planets and stars by turbulent convection, which, coupled with rotation, generates helical flows that can sustain magnetic fields. 
In some cases, differential rotation and velocity shear can be significant and coherent enough to play a crucial role.
In mean-field dynamo theory, when velocity shear and/or differential rotation play a relevant role, the mechanism is called the $\alpha$-$\Omega$ dynamo \citep{Parker_1955}; when it plays a minor role, magnetic fields are generated by helical convection through the $\alpha$ effect, giving an $\alpha^2$ dynamo \citep[e.g.,][]{Chabrier_Kuker_2006, Kapyla_et_al_2009ApJ, Shulyak_2017NatAs, Bushby_et_al_2018AA}.  
Our Sun, where a shear layer known as the tachocline may play a key role in the solar dynamo, might be classified as an $\alpha$–$\Omega$ dynamo. However, the actual dominant mechanism producing the solar magnetic field is still under debate \citep{Vasil_Nature_2024, Zweibel_Nature_2024, Brandenburg_et_al_ApJ_2025}, although different types of dynamo can work together in the same system.

In this study, we focus on the dynamo mechanism of late M dwarfs, which are fast-rotating, fully convective stars that lack a tachocline. Observations indicate that surface differential rotation is minimal in these stars, as most fast-rotating M dwarfs appear to have a nearly solid-body rotation, such as V374 Peg \citep{Donati_et_al_Sci_2006, Morin_et_al_2008MNRAS_a} or GJ 1243 \citep{Davenport_et_al_AJ_2020}. In these stars, the likely absence of significant differential rotation disfavours the operation of an $\alpha$-$\Omega$ dynamo, indicating instead that their fields are sustained by helical convection \citep[e.g.,][]{Chabrier_Kuker_2006,  Kapyla_et_al_2009ApJ, Shulyak_2017NatAs, Bushby_et_al_2018AA}. 
Nevertheless, some degree of differential rotation could also contribute to field generation in fully convective stars, when it remains coherent in time and space \citep{Nigro_et_al_2017, Peera_et_al_2016}. However, an $\alpha^2$ dynamo for fast-rotating M dwarfs is also supported by the stellar X-ray activity-rotation diagram, since X-ray emission is a proxy of dynamo efficiency \citep{Wright_et_al_2011ApJ, Pevtsov_et_al_2003ApJ, Pallavicini_et_al_1981ApJ, Vilhu_Walter_1987ApJ, Fleming_et_al_1995ApJ, Krishnamurthi_et_al_1998ApJ, Magaudda_et_al_2020A&A, Magaudda_et_al_2022}.
Additionally, numerical simulations of fully convective stars indicate a quenching of differential rotation \citep{Browning_2008ApJ}, particularly in the most rapidly rotating models \citep[e.g.][]{Kapyla_2021A&A, Brun_2022ApJ, Bice_Toomre_2023ApJ}. 

Following \citet{Christensen_et_al_2009Natur, Christensen_Aubert_2006}, who show that the magnetic energy density appears to be determined by a scaling law based on the available thermal flux in different astrophysical systems (i.e. Earth, Jupiter, T Tauri stars and M dwarfs) we propose a significant role for the thermal convective flux in the $\alpha^2$ dynamo mechanism in Earth and fully convective M dwarf stars. Although their convective zones differ and may operate in different regimes, the role of heat flux in their dynamos could be similar. Numerical geodynamo simulations have further shown that increased heat flow at the core-mantle boundary is associated with periods of elevated reversal frequency \citet{Glatzmaier_et_al_1999, Valet_et_al_2005, Lay_et_al_2008, Olson_et_al_2010, Driscoll_Olson_2011, Biggin_et_al_2012, Olson_Hagay_2015}.
Considering the idea of a common dynamo mechanism for both Earth and M dwarfs and the central role of thermal flux in geodynamo simulations, it is reasonable to hypothesize that thermal flux may play, in the dynamo of late M dwarfs, a similarly significant role as that observed in the modulation of geomagnetic reversal frequency.

Observational evidence reveals a dichotomy in the magnetic properties of late M dwarfs (spectral types M3/M4 to very late M), which share similar stellar parameters, namely masses below 0.15 M$_\odot$, and rotational periods shorter than four days \citep{Morin_etal_2010MNRAS, Morin_et_al_2011MNRAS}. 
In particular, some M dwarfs have a stable axisymmetric dipole-dominated magnetic field, while other late M stars exhibit considerably stronger variability of their weak global magnetic field \citep{Gastine_et_al_2013, Shulyak_2017NatAs}, often with a significant non-axisymmetric contribution \citep{Kochukhov_2021AA}. 
This dichotomy cannot be explained solely by stellar parameters, leaving its origin unclear. 

Given the central role of thermal flux in $\alpha^2$ dynamos, we developed an MHD shell model for an $\alpha^2$ dynamo and magnetoconvection driven by temperature perturbations. The model captures key statistical features of magnetic reversals and is validated by comparing its results to geomagnetic polarity reversal data, as the Earth's dynamo is the only natural $\alpha^2$ dynamo with a suitable data sample of reversals. We then extend our results to M dwarfs as far as the role of heat flux in magnetic reversals is concerned, and propose that differences in thermal flux efficiency may underlie the observed magnetic dichotomy. Thus, the analogy we propose between the Earth's and M-dwarf dynamos refers exclusively to shared trends and behaviors involving the heat flux, without conflating the specific physical properties of these systems.

We adopted a modified shell model to address two main issues related to the limitation of direct numerical simulations (DNS) and other models: 1. Limited dynamical ranges, as DNS fail to cover the extreme regimes that are instead relevant to astrophysical dynamos; 2. Inadequate statistics for polarity reversals that are crucial to validate an $\alpha^2$ dynamo model.
Regarding the first issue, turbulent magneto-convection in late-type stars is characterized by a very wide range of spatial and temporal scales, with Reynolds numbers Re $\sim 10^{15}$. All the numerical codes adopted for magneto-convection have limited numerical resolution, which is unsuitable for capturing the global multiscale dynamics of this process. 
Regarding the second issue, simplified numerical models such as shell models can reproduce large samples of magnetic reversals with minimal computational cost, providing a useful framework to explore the role of key physical parameters. 
The success of shell models lies in their ability to reproduce, in wide dynamical ranges, the main statistical features of velocity in a turbulent flow with low computational effort \citep{Gledzer_1973, Yamada_Ohkitani_1988PRL, Yamada_Ohkitani_1988PThPh, Ohkitani_Yamada_1989, Kadanoff_et_al_1995PhFl, Gloaguen_et_al_1985, Biferale_2003, Plunian_et_al_2013}, as well as the behavior of passive scalar fields \citep{Ruiz_Nelson_1981PhRvA, Jensen_et_al_1992}. In addition, shell models can reproduce the basic dynamical characteristics of turbulent cascades as observed in magnetohydrodynamic (MHD) simulations \citep{Nigro_2004, Cadavid_et_al_2014ApJ, Nigro_2015} and solar wind plasma \citep{Dominguez_2017, Dominguez_2018, Dominguez_2020}, as well as fractality features like those observed in geomagnetic activity data \citep{Munoz_et_al_2018, Munoz_et_al_2021}. 

Shell models have been proposed to describe thermal convective turbulence \citep{Jensen_et_al_1992, Suzuki_Sadayoshi_1995}, where the velocity advects a passive scalar field. Moreover, \citet{Brandenburg_1992} proposed a shell model to describe hydromagnetic Boussinesq convection adopting real scalar variables, and \citet{Mingshun_Shida_1997} studied the scaling behavior of velocity and temperature in a Boussinesq thermal convection shell model. 
Finally, MHD shell models have shown a turbulent dynamo effect, namely the growth of the magnetic field on the same scales of turbulence starting from a sporadic seed, i.e., small-scale dynamo \citep{Frick_Sokoloff_1998,  Verma_Kumar_2016JTurb}. They are also adopted to describe small-scale turbulence in the context of the large-scale dynamo mechanism consistently with the scale separation approach \citep{Nigro_2011ApJ, Perrone_et_al_2011ApJ, Nigro_2013GApFD}.

In this paper, we study some of the statistical features of a thermally driven magnetoconvection shell model \citep{Nigro_2022}. The model equations are written according to the Boussinesq approximation, and a pitchfork bifurcation term is considered in the largest-scale magnetic field equation to describe magnetic-polarity reversals. Our model captures the key statistical features of the turbulence generated by thermal instability in a simplified but computationally efficient framework. Due to the low computational effort of the model, we can provide a broad sample of numerical simulations by varying the characteristic parameters over a wide range of values that are not yet accessible by DNS. These results offer insight into convective dynamos across regimes relevant to naturally occurring systems that align with our assumptions.  

Our model adopts the Boussinesq approximation and does not include an explicit rotation term. 
Although the Coriolis force can significantly affect the organization of convective flows as observed in global three-dimensional MHD simulations \citep{Brown_et_al_ApJ_2008}, in the present simplified framework, its inclusion would mainly rescale the kinetic energy, without qualitatively altering the nonlinear dynamical regimes or the relative trends in the reversal frequency observed in our simulations. 
Rotational effects are implicitly included in our description through maximally helical convective forcing consistent with low-Rossby-number regimes. Thus, our results are most applicable to fast-rotating systems, where stratification does not dominate the convective dynamics and differential rotation is weak.
We emphasize that our model is highly idealized: it neglects density stratification, which is known to influence the magnetic topology and amplitude of stellar fields \citep[e.g.][]{Gastine_Wicht_AA_2012, Brun_2022ApJ}. Our aim is not to reproduce the full dynamical complexity of M-dwarfs, but rather to explore the role of thermal convection and nonlinear feedbacks on magnetic variability within a simplified $\alpha^2$ dynamo framework. In this sense, our results should be considered as indicative trends that can complement more realistic global simulations. 

In the article, Section \ref{Model} describes the model, Section \ref{results} presents the numerical results, and Section \ref{statistics} examines the statistical results of the magnetic reversals obtained in terms of signed measure and probability distribution functions of persistence times, a persistence time being an interval between two consecutive reversals. In section \ref{heat_flux}, we explore the influence of the Nusselt number on polarity reversals, and in section \ref{conclusions}, we consider possible implications for the dynamo dichotomy in M dwarfs.

\section{Magneto-convection Shell Model}
\label{Model}
The model equations are derived by starting from the extension of a thermally driven convection shell model \citep{Mingshun_Shida_1997} to include the induction equation with the aim of accounting for plasma magnetization effects, which are clearly fundamental in describing the dynamo mechanism. 
This extension builds on previous models, such as those proposed by \citet{Brandenburg_1992} and \citet{Jensen_et_al_1992}, while incorporating terms specifically tailored to reproduce the nonlinear feedback characteristic of $\alpha^2$ dynamos. For additional details on the derivation and rationale behind the model, see Appendix \ref{sec:ApexA}.

The equations of a thermally driven MHD shell model with kinematic viscosity $\nu$, magnetic diffusivity $\eta$, and thermal diffusivity $\chi$, are written for velocity field fluctuations $u_n$, magnetic field fluctuations $b_n$, and temperature fluctuation $\theta_n$ as follows:

\begin{widetext}
\begin{eqnarray}
\hspace*{-0.7cm} \left(\frac{d}{dt} + \nu k_n^2\right) u_n &=&  - \tilde{\alpha} \theta_n + {i} k_n[(u_{n+1}u_{n+2}-b_{n+1}b_{n+2}) \!
- \! \frac{\xi}{2}(u_{n-1}u_{n+1}-b_{n-1}b_{n+1}) \!
- \! \frac{1-\xi}{4}(u_{n-2}u_{n-1}-b_{n-2}b_{n-1})]^* 
\label{eq:GOY_u}
\\
\hspace*{-0.7cm} \left(\frac{d}{dt} + \eta k_n^2\right) b_n  &=&
{i} k_n [(\!1\!\!-\!\xi\!-\!\xi_m\!)(\!u_{n+1}b_{n+2}\!-\!b_{n+1}u_{n+2})
\!+ \! \frac{\xi_m}{2}(\!u_{n-1}b_{n+1}\!-\!b_{n-1}u_{n+1}) \!+ \!
\frac{1-\xi_m}{4}\!(\!u_{n-2}b_{n-1}\!-\!b_{n-2}u_{n-1})]^*
\label{eq:GOY_b}
\\
\hspace*{-0.7cm} \left( \frac{d}{dt} +  \chi k_n^2\right) \theta_n &=&
{i} k_n [\alpha_1 u_{n+1}^*\theta_{n+2}^* + \alpha_2 u_{n+2}^* \theta_{n+1}^* 
+ \beta_1 u_{n-1} \theta_{n+1} - \beta_2 u_{n+1} \theta_{n-1} 
+ \gamma_1 u_{n-1} \theta_{n-2} + \gamma_2 u_{n-2} \theta_{n-1}]^* \!+ \! f_n
\label{eq:Thermal}
\end{eqnarray}
\end{widetext}
%
where $n$ is the shell number ($n=1, \dots, N$, N being the total number of shells considered); $n$ corresponds to the shell of wave vectors with radius $k_n = k_0 2^n$, where $k_0 = \pi/L $ and $L$ is the typical size of the system. Moreover, $ i = \sqrt{-1} $, and $\tilde{\alpha}$ denotes the thermal expansion coefficient, which is proportional to the Rayleigh number Ra; finally, $f_n$ is the forcing that activates convection (see below). The term proportional to $\tilde\alpha$ is the buoyancy term that acts on each shell to convert thermal energy into kinetic energy.

The values of the shell model coefficients $\xi = 1/2$ and $\xi_m = 1/3$ are set to ensure that equation (\ref{eq:GOY_u}) and (\ref{eq:GOY_b}) preserve the quadratic invariants of MHD -- total energy, cross-helicity, and magnetic helicity -- in the ideal case, namely where there is no external forcing or dissipation, and when $\tilde{\alpha} = 0$ \citep{Frick_Sokoloff_1998, Biferale_2003}. These conservation laws are satisfied by the original MHD equations (\ref{velocity}) and (\ref{mag_field}) in the ideal case. Therefore, equations (\ref{eq:GOY_u})--(\ref{eq:GOY_b})--(\ref{eq:Thermal}) coincide with the equations of the MHD Gledzer-Ohkitani-Yamada (GOY) shell model when $\tilde{\alpha} = 0 $ and $\theta_n = 0$ for every $n=1,\dots,N$ \citep{Gledzer_1973, Ohkitani_Yamada_1989}. Whereas $\alpha_1 = \alpha_2 = 1$, $\beta_1 = \beta_2 = 1/2$, and $\gamma_1 = \gamma_2 = - 1/4$ are shell model coefficients whose values are selected by adopting the \citet{Jensen_et_al_1992} coupling model.

The shell model equations (\ref{eq:GOY_u})-(\ref{eq:GOY_b})-(\ref{eq:Thermal}) are obtained by adopting the Boussinesq approximation, assuming an incompressible plasma and neglecting density stratification. This simplification allows for a broader exploration of parameter space, including extreme Reynolds and Rayleigh numbers, but limits its direct applicability to systems where stratification does not significantly affect convection. These approximations prioritize computational efficiency over detailed physical realism, enabling the statistical study of magnetic reversals across a wide range of conditions.

In addition, the shell model equations are written under the assumption of local interaction between the shells \citep[e.g.][]{Gledzer_1973, Ohkitani_Yamada_1989}.
In fact, equations (\ref{eq:GOY_u})-(\ref{eq:GOY_b})-(\ref{eq:Thermal}) do not capture a key ingredient of the dynamo mechanism, which is instead highlighted in mean-field electrodynamics (MFE). This aspect involves the interaction of dynamics on different scales, which is described by the scale separation approach adopted in MFE \citep{Parker_1955, steenbeck_1966, Moffatt_1978, Krause_Radler_1980}. 
The success of MFE theory is due to the magnetic field characteristic of revealing itself through the tendency to self-organize into coherent structures with spatial and temporal scales larger than the underlying turbulence \citep{Nigro_et_al_2017, Peera_et_al_2016, Nigro_2019NCimC}. Therefore, considering the success of this theory, we found it important to incorporate non-local interaction among scales into the model.
Therefore, according to the MFE and adopting the quasi-linear approximation, an electromotive force $\epsilon = \alpha\;{b_1}$ has to drive the dynamo mechanism in the equation of the large-scale magnetic field ${ b_1}$, we adopt the following quenching function for $\alpha$:
\begin{equation}
\label{eq:alpha1}
\alpha=\mu \left(1- \frac{{b}_1^2}{B^2_{0}}\right) \;.
\end{equation}
This term accounts for the nonlinear back-reaction of the Lorentz force, ensuring that the system saturates at equipartition energy levels. Indeed, $B_0$ is the equipartition magnetic field amplitude \footnote{An equipartition regime is achieved when the energy per unit volume of the mean magnetic field is comparable to the energy of the underlying turbulent fluid motion.}. This quenching mechanism prevents the unphysical growth of the magnetic field and is in agreement with the mean-field dynamo theory \citep{Brandenburg_Subramanian_2005PhR, Charbonneau_2010LRSP}. Basically, the cubic nonlinearity of the term $\alpha b_1$ in the induction equation ensures that the electromotive force is an odd function of $b_1$, allowing field reversals, in agreement with the arguments raised by \citet{Field_Blackman_2002ApJ}. This formulation produces a supercritical pitchfork bifurcation -- a well-studied route to polarity reversals in both hydrodynamic and MHD systems \citep[e.g.,][]{Tobias_1996, Sorriso_et_al_2007PEPI, Benzi_Pinton_2010PhRvL, Benzi_Pinton_2011IJBC, Nigro_Carbone_2011}. 
Furthermore, $\mu$ is a constant related to the kinetic helicity of flow motions, which is at its maximum value, as defined by the construction of our shell model, in agreement with the very low Rossby number regime. Therefore, we set $\mu = 1$ as the typical value of normalized kinetic helicity is $H_k \approx 1$. In addition, we assume that an equipartition level is achieved in the saturation regime. 

Therefore, the resulting equation for the magnetic field in the first shell is the following:
\begin{equation}
\frac{d b_1}{dt}= - \eta k_1^2 b_1 +
{ i} \frac{k_1}{6}\; (u_{2}^*b_{3}-b_{2}^*u_{3})+ \mu b_1 \left( 1- \frac{b_1^2}{B_0^2}\right) \ .
\label{eq:b1}
\end{equation}
It is important to note that in this model we do not describe the magnetic field topology on large scales, as the adopted shell model formulation does not allow for such a representation. Consequently, the magnetic field $b_1$, corresponding to the first shell, essentially represents the Fourier components within the largest scale range, without distinguishing between dipolar and toroidal modes.

In equation \ref{eq:b1}, the cubic term in $b_1$ describes a supercritical pitchfork bifurcation, where $b_1=0$ is an unstable equilibrium, while $b_1 = \pm B_0$ are two stable equilibria (two fixed points in phase space). The pitchfork bifurcation has been successfully adopted in a modified hydrodynamic shell model to describe sudden flow reversals \citep{Benzi_2005} as those observed in the large-scale circulation of turbulent Rayleigh-B{\'e}nard convection or the wind acceleration in the Earth's atmosphere.
Recently, this kind of one-dimensional unstable manifold arising from a pitchfork bifurcation has been designed to describe magnetic-polarity reversals in a modified shell model \citep{Nigro_Carbone_2011}, and it turns out to be very similar to the term adopted by \citet{Benzi_Pinton_2010PhRvL}. 
Equation (\ref{eq:b1}) substitutes the equation for the magnetic field fluctuation in the first shell ($n=1$) in the system (\ref{eq:GOY_u})--(\ref{eq:GOY_b})--(\ref{eq:Thermal}). The equation (\ref{eq:b1}) thus has to be numerically resolved jointly with all the other shell modes' equations. 
The dynamics of the large-scale magnetic field (\ref{eq:b1}) is coupled with the field fluctuations of neighboring scales, specifically shells $n=2$ and $n=3$, by the term ${ i} \frac{k_1}{6}\; (u_{2}^*b_{3}-b_{2}^*u_{3})$. The action of the small-scale dynamics on the large-scale magnetic field is described by the transport coefficient $\alpha$ given by equation (\ref{eq:alpha1}), which produces a supercritical Pitchfork bifurcation. 
Pitchfork bifurcation, well-known in bifurcation theory, has been observed in several models studying the dynamo transition.
Indeed, this bifurcation clearly appears in low-order dynamo systems obtained by mode truncation \citep[e.g.][]{Schmalz_Stix_1991, Verma_et_al_2008PhRvE, Rohit_Pankaj_2017}, in untruncated dynamo simulations with dipole symmetry \citep{Tobias_1996, Tobias_1997, Passos_et_al_2012}, 
as well as in direct numerical simulations of the dynamo transition as those performed by \citet{Dubrulle_et_al_2007NJPh} and \citet{Yadav_et_al_2010} in a Taylor-Green flow. Basically, a cubic term like (\ref{eq:alpha1}) has been widely present in the dynamo model literature since the work of \citet{Stix_1972, Meinel_Brandenburg_1990}.

The external forcing term causes temperature fluctuations, driving the plasma flow to become unstable. This forcing is the solution of the following Langevin stochastic equation, namely 
\begin{align}
&df_n/dt = -f_n/\tau_c + \gamma \;\;\;\;\;\; {\rm when} \;\;\; n=1,2,3;
\label{langevin}
\\
&{\rm whereas} \;\; f_n = 0 \;\;\;\;\;\; {\rm  when} \;\;\; n=4,5,...,N;
\nonumber
\end{align}
where $\tau_c$ is the correlation time that we set $\tau_c=1$; $\gamma(t)$ is a Gaussian stochastic process $\delta$-correlated in time $\langle \gamma(t) \gamma(t^\prime) \rangle = \sigma \delta(t-t^\prime)$, with the constant $\sigma = 1$. The physical motivation of this choice is to represent a forcing term that mimics the effect of large-scale thermal fluctuations, which are not purely random (not white noise) but have a short-term memory associated with the correlation times of the convective system. This forcing corresponds to temperature fluctuations that develop on spatial and temporal scales comparable to those of the convective region.

It is important to note that the final model equations no longer conserve the MHD quadratic invariants due to the $\alpha$ term (in addition to the forcing and dissipative terms). Specifically, the magnetic helicity is not constant.
Indeed, our model is not actually `closed' because we do not describe the small-scale dynamics that should, in principle, be solved consistently with the large-scale fields, whereas a prescribed $\alpha$ term is adopted. 
In this regard, our scenario can be considered in line with the models that allow magnetic helicity fluxes \citep[e.g.][]{Hubbard_Brandenburg_2010GApFD, Kapyla_et_al_2010A&A, Del_Sordo_et_al_2013MNRAS, Brandenburg_2018JPlPh, Cattaneo_Bodo_Tobias_2020_2020JPlPh}, and it is free from any catastrophic $\alpha$ quenching (see \citet{Tobias_2021JFM} and references therein).
In summary, the model offers a computationally efficient framework for studying $\alpha^2$ dynamo mechanisms, enabling access to extreme parameter regimes (e.g., high Reynolds and Rayleigh numbers) that are inaccessible to DNS. By incorporating both local and nonlocal interactions, it bridges the gap between simplified shell models and the more complex dynamics of naturally occurring dynamos. The equations, along with the modifications derived for non-local interactions, are solved simultaneously to capture the essential features of the magnetoconvective system.
\section{Numerical Simulations}
\label{results} 
The model equations (\ref{eq:GOY_u}), (\ref{eq:GOY_b}) and (\ref{eq:Thermal}) are written in dimensionless units. We adopt the freefall velocity $U = \sqrt{ \hat{\alpha} g L \Delta T}$ as a velocity unit and measure the magnetic field in terms of $ B_0 = U$, which is the typical magnetic field obtained when the freefall velocity $U$ is equal to the Alfv\'en speed. The temperature is measured in terms of $\Delta T$, the length in units of $L$, i.e. the characteristic length of the system, and the time in units of freefall time $( L/ \hat{\alpha} g \Delta T)^{1/2}$ (please, see Appendix \ref{sec:ApexA} for the meaning of the above quantities). 

A fourth-order Runge-Kutta numerical scheme is adopted to solve the model equations, considering 19 shells $N=19$ to capture a wide range of scales. The saturation amplitude of the large-scale magnetic field is set to $B_0=1$, in dimensionless units, which corresponds to the energy equipartition scenario. The dimensionless correlation time of the forcing is chosen as $\tau_c=1$. 

To explore the parameter space, we vary key quantities within physically relevant ranges:
\begin{itemize}
    \item Dimensionless thermal convection parameter $\tilde{\alpha} \in [0.1;10]$, proportional to the Rayleigh number Ra.
    \item Dimensionless kinematic viscosity $\nu \in [10^{-1};10^{-6}]$, affecting the hydrodynamic Reynolds number (Re).
    \item Dimensionless magnetic diffusivity $\eta \in [10^{-2};10^{-6}]$, linked to the magnetic Reynolds number (Rm).
    \item Dimensionless thermal diffusivity $\chi \in [10^{-1};10^{-9}]$, which influences Ra and convective dynamics.
\end{itemize}
In fact, $\tilde{\alpha}$ is proportional to the Rayleigh number Ra $= {\tilde{\alpha}}{\theta_0}L^3/(\nu \chi)$, where $\theta_0 =  \sqrt \langle {\sum_{n=1}^N \theta_n^2}\rangle_{t}$.
The hydrodynamic Reynolds number is Re$=Lu_0/\nu$ and the magnetic Reynolds number is Rm$=Lu_0/\eta$, where $u_0 = \sqrt \langle {\sum_{n=1}^N u_n^2} \rangle_{t}$ and $b_0 = \sqrt \langle {\sum_{n=1}^N b_n^2}\rangle_{t}$. 
The typical dimensionless values in our simulations are $\theta_0 \approx u_0 \approx b_0 \approx 1$.

\begin{figure*}
\centering 
\subfloat[Considering the same interval of time, the number of the magnetic field $b_1(t)$ polarity reversals increases for an increasing Rayleigh number (decreasing diffusivity $\chi$).]{%
  \includegraphics[width=0.99\textwidth]{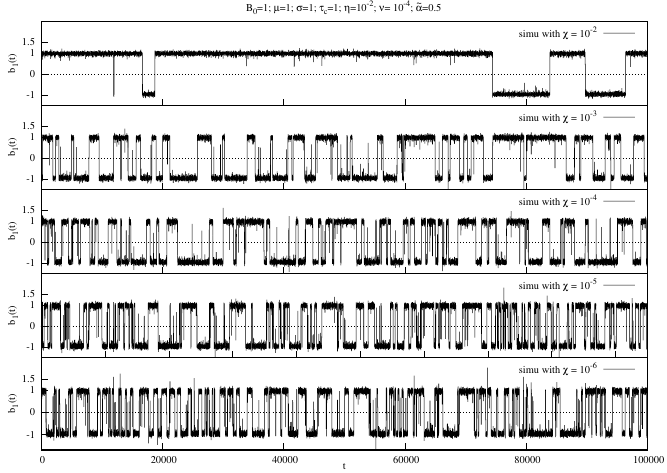}%
  \label{fig:vs_chi_dyn}}\qquad
\subfloat[The tendency to develop shorter persistence times for an increasing Rayleigh number is shown by the PDFs on the left and sign-singularity in the magnetic reversals as revealed by Partition Functions $\zeta$, on the right.  
 On the left, a linear fit is applied to the PDF wings with the slope indicated in the corresponding area of the fitting line. Black triangles represent the PDF inferred from the CK95 database for the Earth's polarity reversals, which are being introduced for comparison.  
On the right, the linear fit is made in a suitable inertial range for each partition function depicted with slopes, respectively, 
$-0.49 \pm 0.02$ for the simulation obtained by setting $\chi = 10^{-2}$, 
$-0.50 \pm 0.01$ when $\chi = 10^{-4}$, and finally $-0.53 \pm 0.02$ when $\chi = 10^{-6}$.
We also computed the partition function on the CK95 dataset with slope and error, as shown in the plot.]
{%
  \includegraphics[width=0.99\textwidth]{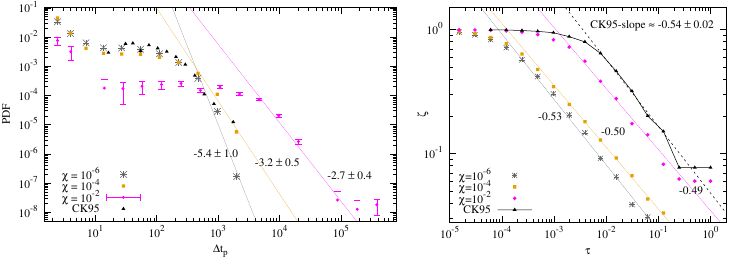}%
  \label{fig:vs_chi}%
}
\caption{Sensitivity of the system with respect to the thermal diffusivity $\chi$ (i.e., Rayleigh number $\textrm{Ra} = {\tilde{\alpha}}{\theta_0}L^3/(\nu \chi))$, keeping constant the other parameters; in particular $\tilde{\alpha} =0.5 $, $\eta = 10^{-2}$, and $\nu = 10^{-4}$.}
\label{fig:whole_vs_chi}
\end{figure*}

The simulations reveal a clear dependence of the magnetic polarity reversal frequency on key parameters:
\begin{itemize}
    \item Increasing $\tilde{\alpha}$ (or Ra) results in more frequent reversals within the same time interval (Fig.~\ref{fig:vs_alpha_dyn}, see also \citep{Nigro_2022}).
    \item Reducing $\nu$ or $\eta$ enhances reversal frequency, reflecting higher Re and Rm values (Figs.~\ref{fig:vs_nu_dyn}, \ref{fig:vs_eta_dyn}).
    \item Lowering $\chi$, which increases Ra, similarly increases reversal frequency (Figs.~\ref{fig:vs_chi_dyn}, \ref{fig:vs_chi}).
\end{itemize}
These results highlight the system's sensitivity to both convective and dissipative properties. Specifically, variations in $\tilde{\alpha}$ produce more pronounced changes in reversal dynamics compared to equivalent changes in diffusivity coefficients.

%
%
%
 \subsection{PDFs and Sign-Singular Measure} 
 \label{statistics}
 In this section, we validate our model by comparing the statistical features of the magnetic polarity reversal calculated from our numerical results with those obtained by analyzing paleomagnetic records \citep[e.g.][]{CK95, Constable_2000PEPI}.

 Considering simulations lasting for a much longer time than those depicted in Figures \ref{fig:vs_chi_dyn}, \ref{fig:vs_alpha_dyn}, \ref{fig:vs_nu_dyn} and \ref{fig:vs_eta_dyn} (that means $\sim 5 \times 10^6$ unit time), we can obtain a suitable sample of persistence times that can be statistically analyzed in each case considered. Persistence time is an interval of time during which the large-scale magnetic field maintains the same polarity, i.e., the same sign.  A rich sample of persistence time for each simulation allows us to statistically analyze these, computing their probability distribution functions (PDFs) and the signed measure. 

Studying the PDFs of persistence times enables us to investigate the nature of magnetic reversals in our system. In particular, the slope of the PDF tail reveals the eventual presence of memory in the process. Indeed, an exponential decay is characteristic of a Poisson process, i.e., a random process. In contrast, a power law tail suggests that the magnetic field inverses with some level of memory, where past events affect the future probability of inversion and the system ``recalls'' a state for a certain time before moving to another state.
In addition, the width of the PDF provides information on the relative stability of the magnetic polarity. A broader PDF, with significant probability at long persistence times, indicates that the system tends to maintain the same polarity for extended periods, suggesting a robust large-scale magnetic configuration. Conversely, a narrower PDF, decaying more rapidly, corresponds to more frequent polarity changes, and thus a less stable magnetic field. 

The PDFs of persistence times obtained in our model show power-law tails, indicating a deviation from a Poisson process, although small. This deviation may be due to long-range correlations that usually characterize processes that keep memory effects in their occurrence. This result agrees with the paradigm proposed by \citet{Sorriso_et_al_2007PEPI}, who found a clustering of the Earth's magnetic field reversals. Although the exact distribution of persistence times is debated in the geodynamo community \citep[e.g.][]{Merrill_et_al, Gubbins_2008Natur, Olson_et_al_2010, Shcherbakov_Fabian_2012JGRB}, our model produces polarity reversals with PDFs that closely match paleomagnetic data (see Fig.~\ref{fig:vs_chi}).

In addition, our statistical analysis shows a tendency for the system to develop longer persistence times and larger PDF, as the dissipative coefficients increase or the thermal convection coefficient $\tilde\alpha$ decreases, indicating that weaker convective driving corresponds to more stable magnetic configurations. This behavior is clarified by the PDFs of the persistence times that extend toward longer periods, showing thus decreasing slopes when dissipative coefficients increase, or $\tilde\alpha$ decreases (see Table \ref{tab:slopes} or/and Figs.~\ref{fig:vs_alpha}, \ref{fig:vs_nu}, \ref{fig:vs_eta} and \ref{fig:vs_chi}).

However, the system's sensitivity to the different coefficients is not the same. In fact, the PDFs of the persistence times have a much sharper change when we vary $\tilde{\alpha}$ compared to the changes they undergo when we vary the dissipative coefficients. Comparing Fig.~\ref{fig:whole_vs_alpha} with Figs.~\ref{fig:whole_vs_chi}, \ref{fig:whole_vs_nu}, and \ref{fig:whole_vs_eta}, we can observe that the system has a similar change in the frequency of reversals by varying the value of the dimensionless dissipative coefficient in steps of ten units, while dimensionless $\tilde{\alpha}$ has to vary in steps of $0.5$ to produce a similar change (see also Table \ref{tab:slopes}).
%
%
\begin{deluxetable}{cccccc}
  \tablenum{1}
  \tablecaption{Slopes of the persistence times PDFs and the cancellation exponent $\kappa$ in 4 simulation models. \label{tab:slopes}}
  \tablewidth{0pt}
  \tablehead{
    \colhead{$\tilde{\alpha}$} & \colhead{$\chi$} & \colhead{$\eta$} & \colhead{$\nu$} & \colhead{PDF slope} & \colhead{$\kappa$} 
  }
  \startdata
0.5 & $10^{-2}$ & $10^{-2}$ & $10^{-4}$ & $-2.7\pm0.4$ & $0.49\pm0.02$\\
0.5 & $10^{-4}$ & $10^{-2}$ & $10^{-4}$ & $-3.2\pm0.5$ & $0.50\pm0.01$\\
0.5 & $10^{-6}$ & $10^{-2}$ & $10^{-4}$ & $-5.4\pm1.0$ & $0.53\pm0.02$\\
\hline
0.5 & $10^{-2}$ & $10^{-2}$ & $10^{-2}$ & $-2.1\pm0.2$ & $0.51\pm0.30$\\
1.0 & $10^{-2}$ & $10^{-2}$ & $10^{-2}$ & $-3.8\pm0.6$ & $0.53\pm0.02$\\
1.5 & $10^{-2}$ & $10^{-2}$ & $10^{-2}$ & $-4.8\pm1.0$ & $0.55\pm0.02$\\
\hline
0.5 & $10^{-5}$ & $10^{-2}$ & $10^{-5}$ & $-3.3\pm0.9$ & $0.49\pm0.02$\\
0.5 & $10^{-5}$ & $10^{-4}$ & $10^{-5}$ & $-4.3\pm0.5$ & $0.50\pm0.02$\\
0.5 & $10^{-5}$ & $10^{-6}$ & $10^{-5}$ & $-5.2\pm0.8$ & $0.52\pm0.01$\\
\hline
0.5 & $10^{-2}$ & $10^{-4}$ & $10^{-2}$ & $-2.1\pm0.2$ & $0.45\pm0.06$\\
0.5 & $10^{-2}$ & $10^{-4}$ & $10^{-4}$ & $-3.8\pm0.6$ & $0.47\pm0.02$\\
0.5 & $10^{-2}$ & $10^{-4}$ & $10^{-6}$ & $-4.8\pm1.0$ & $0.52\pm0.01$
\enddata
\tablecomments{In each of the 4 simulation models, we changed the value of only one parameter. The 4 different simulation models are separated in the table by a horizontal line. In each model, the PDF slope and $\kappa$ increase in the absolute value as the corresponding dissipative coefficient decreases or, as in the second model, the thermal expansion coefficient $\tilde{\alpha}$ increases.}
\end{deluxetable}

It is also interesting to note that the changes in the shape of the PDFs are not linear as the dissipative coefficients decrease or $\tilde{\alpha}$ increases. 
In particular, when $\chi$ decreases from $10^{-2}$ down to $10^{-4}$, the mean persistence time is reduced more strongly than that observed when $\chi$ decreases from $10^{-4}$ down to $10^{-6}$ (see Fig.~\ref{fig:whole_vs_chi}). 
The same sensitivity of the system is observed in the reduction of $\nu$ and $\eta$ concerning the same range of variation examined for $\chi$ (Figs. \ref{fig:whole_vs_nu} and \ref{fig:whole_vs_eta}).
The nonlinear sensitivity of the system is also observed for the parameter $\tilde{\alpha}$: the reduction of persistence time average when $\tilde{\alpha}$ increases from $0.5$ up to $1$ is greater than that detected for $\tilde{\alpha}$ increasing from $1$ up to $1.5$ (Fig.~\ref{fig:whole_vs_alpha}). 
This confirms that the system does not respond uniformly to the parameter variations within the examined range. 

Another statistical quantity we considered is the signed measure that quantifies the intermittency of magnetic fluctuations that reverse polarity. This measure is generally considered when examining a quantity with an alternating sign \citep[e.g.,][]{Ott_et_al_1992, Cadavid_et_al_1994ApJ,  Cattaneo_Tobia_2005PhFl, Martin_et_al_2013PhRvE, Halmos_2018, Sorriso_et_al_2019FrP, LEO}. In fact, probability is a positively defined measure, whereas a signed measure is a function that can take both positive and negative values (it is bounded between 1 and $-1$) and can be interpreted as the difference between two probability measures, one for positive values and the other for negative values of the same field, respectively.  
In particular, we built a signed measure as follows. Starting from the field $b_1(t)$ in the total interval $\Omega_{tot}$, we can define a hierarchy of disjoint subsets $\{\Omega_i\}_{i \in \mathbb{N}}$ of intervals of equal size $\tau$ (normalized to the entire simulation time $\Omega_{tot}$) covering the total interval $\Omega_{tot}$. For each selected time scale $\tau$, a subset $\{\Omega_i\}$ is defined that covers $\Omega_{tot}$. 
We can define for each subset of intervals $\{\Omega_i\}$ the measure $\mu \big(  \bigcup_{i=1}  \Omega_i \big)= \sum_{i=1} \mu({\Omega_i})$, where: 
\begin{equation}
\label{mu_i}
 \mu (\Omega_i, \tau)= \frac{\int_{ \Omega_i} b_1(t) dt}{ \int_{\Omega_{tot}} |b_1(t)| dt } .
\end{equation}
Since our field $b_1(t)$ changes sign by switching between values $\sim +1$ and $\sim -1$, possible cancellations between values of opposite signs can be present in the sum of equation (\ref{mu_i}). 
In principle, the smaller the selected scale $\Delta t_i$, the fewer cancellations should be present. This characteristic could hold continuously for the decreasing selected scale.  
In particular, $\mu$ is {\it sign-singular} if, for any interval $\Omega_i$, no matter how small, there exists a sub-interval $\tilde{\Omega}_i \subset \Omega_i$ such that $\mu({\Omega_i})$ has opposite sign to $\mu(\tilde{\Omega}_i)$. Namely, $\mu$ is sign-singular when $b_1$ changes the sign everywhere on an arbitrary fine scale. 

The rapid change from positive to negative values is a typical feature of a magnetic dynamo. This effect becomes more extreme as the Reynolds numbers increase, approaching a sign singularity when Rm diverges as shown by \citet{Ott_et_al_1992}. 
A way to measure this effect is to calculate the cancellation exponent $\kappa$, which satisfies the following scaling law for the partition function:
\begin{equation}
\zeta(\tau) = \sum_{\Omega_i} | \mu(\Omega_i, \tau)| \sim \tau^{-\kappa} 
\label{kappa}
\end{equation}
where $\kappa$ provides a measure of cancellation efficiency in the range of scales where the scaling law (\ref{kappa}) holds. 
In fact, the higher the value of $\kappa$, the more efficient the cancellation will be. A higher $\kappa$ corresponds to stronger intermittency and more fragmented polarity structures, while lower values indicate a smoother and more coherent field.

As expected, nontrivial cancellation exponents emerge in our model when we compute the partition function $\zeta$ in different simulations obtained by increasing Rayleigh and Reynolds numbers (see Figs. \ref{fig:vs_chi}, \ref{fig:vs_alpha}, \ref{fig:vs_nu} and \ref{fig:vs_eta}).
In particular, $\kappa$ increases with increasing Rayleigh and Reynolds numbers, thus approaching a sign singularity (see Table \ref{tab:slopes}). 
In addition, it is interesting to note that for decreasing scales $\tau$, the measure becomes smoother, and when this is smooth enough, $\zeta$ approaches its saturation value, i.e. 1. The cutoff scale where this saturation occurs represents the scale under which the rapid variations of $b_1$ from positive to negative values end. 
Diffusive processes, namely viscosity and thermal and magnetic diffusivity, usually determine the cutoff scale \citep{Du_et_al_1994} as they smooth out the field $b_1(t)$. In particular, the cut-off scale is smaller for decreasing viscosity and/or diffusivity, hence for larger Rayleigh and Reynolds numbers.

We compare the partition function $\zeta$ computed in some of the simulations made with the one computed on the CK95 dataset, which collects the polarity inversions of the Earth's magnetic field \citep{CK95}. For the latter case, the cutoff scale turns out to be equal to $\tau_{sat} \sim 0.001 \Omega_{tot}$, which corresponds to $\sim 0.07$ Myrs, which is not far from the Milankovitch orbital frequencies \citep{Milankovictch_1969ciip_book}. 
This saturation time can provide information regarding a characteristic process occurring in the Earth's outer core, which is the source of the geomagnetic field. 
 
We also note that the signed measure of the large-scale magnetic field $b_1(t)$ computed in our study shows a slight deviation from a purely stochastic process as the scaling exponent satisfies $\kappa \gtrsim 0.5$ (strictly $\kappa=0.5$ for a random field). 
This weak deviation suggests the presence of existing correlations among reversal events in the geodynamo history within its random nature \citep{Nigro_Carbone_2011}. Our results agree with the clustering found in the paleomagnetic data \citep{Jonkers_2003PEPI}. In fact, \citet{Olson_et_al_2014E&PSL} claimed that reversals are nearly periodic in several portions of the Mesozoic sequence, and \citet{Ryan_Sarson_2008EL} sustained a geodynamo at the edge of chaos through intermittent switching between periodic and chaotic states \citep{Ryan_Sarson_2011PEPI}.
Unfortunately, speculation is still not possible for $\alpha^2$ dynamos other than the geodynamo due to the lack of suitable datasets.

Although the exact distribution of persistence times is still debated in the geodynamo community, the PDFs and signed measures obtained here closely match those derived from paleomagnetic data \citep[e.g.][]{CK95, Constable_2000PEPI, Sorriso_et_al_2007PEPI}. This agreement indicates that our model reproduces the main statistical features of geomagnetic reversals, capturing the essential dynamics of an $\alpha^2$ dynamo, thus providing a valuable framework for studying polarity reversals in both terrestrial and stellar dynamos. 

However, it is important to note that while both the geodynamo and the dynamo of fully convective stars can be classified as a $\alpha^2$ dynamo and can share similar dynamical behaviour and statistical properties, geomagnetic reversals occur over longer timescales than the observed stellar magnetic variability. The governing equations of these systems are indeed solved in their dimensionless form. Therefore, to obtain the characteristic times for each system, we must rescale the numerical results using the appropriate values of the physical parameters. Consequently, our analogy is qualitative rather than quantitative, and the differences in characteristic times among the magnetic fields of late M dwarfs and planets such as Earth may reflect differences in the values of the physical parameters (e.g., magnetic diffusivity) rather than in fundamental dynamical processes.
%
\begin{figure*}
   \includegraphics[width=20cm]{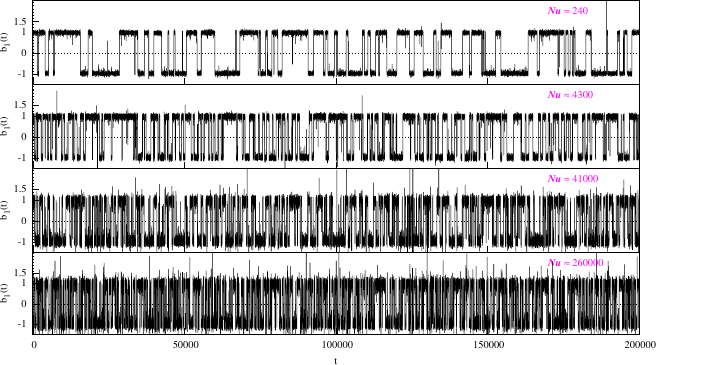}
    \caption{Simulations with higher $Nu$ tend to develop much more reversals than those with lower $Nu$.}
    \label{fig:vs_Nu}
\end{figure*}
\begin{figure} 
   \includegraphics[width=\columnwidth]{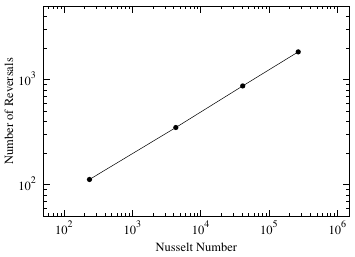}
    \caption{The number of magnetic reversals increases with $Nu$. This trend is depicted here considering the four simulations of Fig.~\ref{fig:vs_Nu}, for which $Nu$ increases from the top to the bottom panel.}
    \label{fig:number_vs_Nu}
\end{figure}
%

\subsection{The role of convective heat transport in the magnetic reversal frequency} 
\label{heat_flux}
In Rayleigh-B{\'e}nard (RB) convection, namely a paradigm of a fluid layer heated from the bottom and cooled at the top, the Nusselt number $Nu$ is introduced to assess the transfer of convective heat within the fluid from the bottom to the top boundary \citep{Ozisik_1985, White_1988}.
Specifically, $Nu$ is a dimensionless measure of the convective heat transport expressed in terms of the ratio of the total heat flux (convective and conductive) and the conductive heat flux \citep{Ahlers_et_al_2009, Chilla_et_al_2012, Verma_book, Pandey_et_al_2021}. Considering $v_z$ as the velocity component along the gravity acceleration, i.e., along the $z$ axis, $u$ as the turbulent vertical velocity and $\Theta$ the temperature fluctuation, we can write, accordingly to \citet{Benzi_et_al_1998}, the following equation
\begin{align}
Nu = \frac{\langle  v_z  T \rangle - \chi \frac{ \partial{\langle T \rangle} }{ \partial z}  }{  \chi \frac{\Delta T}{L} }
= \frac{\langle u \Theta \rangle}{ \chi \frac{\Delta T}{L} } \ ,
\label{eq:Nu1}
\end{align}

where the brackets represent the space and time averages. 
In our model, the velocity is normalized to the free-fall velocity $U = \sqrt{\alpha g L \Delta T} = \sqrt{\nu \chi \textrm{Ra} }/L $, and the temperature to $\Delta T$, $Nu$ can thus be expressed as:
\begin{align}
Nu = \sqrt{\textrm{Ra Pr}} \; \langle | \sum_{n = 1}^N  [{ u_n(t) {\theta_n (t)}^* + {u_n(t)}^* \theta_n (t) }] | \rangle_t \ ,
\label{eq:Nu2}
\end{align}
where Pr is the Prandtl number and the brackets $\langle \rangle_t$ indicate a time average. 

It is interesting to know that in the RB paradigm of a non-magnetized fluid, purely hydrodynamic studies \citep[i.e.][]{Sun_et_al_2005, Xi_et_al_2008, Weiss_Ahlers_2011, van_der_Poel_2011, Xi_et_al_2016} have shown that a temporary increase of the instantaneous $Nu$ number characterizes reversals of large-scale circulation. The instantaneous $Nu$ number can be easily obtained from equation (\ref{eq:Nu2}) without making the average in time. 
The latter-mentioned articles are particularly interesting in our context if we consider the similarities between large-scale flow reversals and magnetic field reversals in the Kolmogorov flow and magnetic dynamo \citep{Gallet_2012}. In particular, \citet{Nigro_2022} shows that a rapid increase in the convective heat flux, i.e., a momentary increase in the instantaneous $Nu$ number, can cause the onset of a magnetic polarity reversal, supporting the idea that the role of convective heat transport may be critical for the occurrence of a magnetic reversal. 

In this article, we further support the idea of a probable key role of the convective heat flux by considering the average global $Nu$ in different simulations we obtained by varying the model parameters. In detail, higher values of $Nu$ characterize those simulations that have more frequent magnetic reversals and, thus, significantly higher magnetic variability (Fig.~\ref{fig:vs_Nu}). This result agrees with geodynamo models and simulations, in which core heat flow is positively correlated with the frequency of polarity reversals and dipole variability \citep[e.g.][]{McFadden_Merrill_1984JGR, Glatzmaier_et_al_1999, Valet_et_al_2005, Lay_et_al_2008, Olson_et_al_2010, Driscoll_Olson_2011, Biggin_et_al_2012, Olson_Hagay_2015}.
On the other hand, Earth, Jupiter, young contracting stars, and rapidly rotating low-mass stars have magnetic fields that satisfy a common empirical scaling law that sets the level of magnetic energy with convected energy flux \citep{Christensen_et_al_2009Natur}. This shared scaling has also been inferred from dynamo simulations tailored to model the Earth's dynamo \citep{Christensen_Aubert_2006}. The remarkable implication of this scaling is that a similar dynamo mechanism is likely to operate in the systems mentioned above, despite vast differences in their physical conditions. 
However, it is necessary to emphasize that it remains unclear how different convective systems (i.e., an almost incompressible liquid metal for the Earth and compressible plasmas for those stars) may operate similarly to produce an analogous dynamo mechanism, precisely an $\alpha^2$ dynamo. 
Based on the quoted scaling law, we assume that the magnetism of both Earth and fast-rotating fully convective stars is regulated by a similar $\alpha^2$ dynamo mechanism. Accordingly, our results can also describe the main dynamic features of the large-scale magnetic field of fast-rotating fully convective stars under the assumption of a density stratification that does not affect significantly convection. This consideration and our numerical results can allow us to propose the idea of a likely critical role of convective heat transfer in the magnetic variability of an $\alpha^2$ dynamo. Indeed, we retain that convective heat transport efficiency can influence the time variability in fully convective stars, which thus are prone to change magnetic polarity more frequently when heat flux is substantially higher. Therefore, the magnetic dichotomy observed in late M dwarfs may be explained in terms of the efficiency of convective heat transfer. 

Notably, considering the same length of time, the number of magnetic reversals increases with $Nu$ as Fig.~\ref{fig:vs_Nu} and Fig.~\ref{fig:number_vs_Nu} show for four simulations where Re has almost the same value. In these simulations, precisely $Nu$ increases because $\chi$ decreases and when the thermal diffusivity coefficient decreases, as a consequence, the velocity amplitudes increase even if the viscosity $\nu$ is kept constant. This increase in velocity amplitudes means that Re does not remain strictly constant but weakly increases with $\chi$. In particular, for the simulations considered in Fig.~\ref{fig:vs_Nu}, Re weakly increases from the top to the bottom panel, starting from Re $\lesssim 2000$ to $\gtrsim 2000$, while correspondingly $Nu$ has a sharp increase.

In this context, considering the specific case of the simulations considered in Fig.~\ref{fig:vs_Nu}, it is important to note that high levels of turbulence might become, in some cases, ineffective for convective heat transport, since what really matters for convective heat transport is the correlation between the temperature and the velocity field. This situation arises in some studies of fluid convection, where uncorrelated temperature and velocity fields very likely coexist within particular states of turbulence \citep[e.g.,][]{Pandey_Sreenivasan_EPL_2021}. Our result, depicted in Fig.~\ref{fig:vs_Nu}, basically means that an increased level of turbulence does not always translate into more efficient heat transport.
Notably, the simulations whose PDFs of persistence times do not extremely change when $\nu$ decreases from the value $10^{-4}$ to $10^{-6}$ (see Fig.~\ref{fig:vs_nu}) are precisely associated with the scenario where enhanced turbulence levels do not substantially increase convective heat transport. Nevertheless, in most of our simulations, we find turbulence states where the Nusselt number value, i.e. the correlation between temperature and velocity field fluctuations, is related to the turbulence level. Specifically, $Nu$ is generally positively correlated with the Reynolds numbers Re and Rm. 
Indeed, the level of turbulence is significant for the occurrence of magnetic reversals, as Figures \ref{fig:vs_nu_dyn} and \ref{fig:vs_eta_dyn} show. In other words, generally, the higher the level of turbulence in the magnetofluid, the greater the number of magnetic reversals.

\section{Discussion and Conclusions}
\label{conclusions}
The magnetic dynamo is the mechanism responsible for the generation and evolution of astrophysical magnetic fields. Different natural dynamo systems exhibit a remarkable ability to switch their polarity and produce magnetic reversals, a phenomenon that is still not entirely understood. The source of the dynamo mechanism is magneto-convection coupled with rotation, which occurs in planets like Earth and Jupiter, as well as in very late-type stars considered fully convective. In these systems, the $\alpha^2$ dynamo can be considered a common mechanism for generating magnetic fields, as the possible shear velocities do not appear to remain coherent in space and over time.  

A magneto-convection shell model for $\alpha^2$ dynamos is developed according to the Boussinesq approximation, where a buoyancy term has been included to consider the interaction of temperature fluctuations with eddies of the same size. The model discussed here is a highly idealized description of magneto-convection, which is a complex nonlinear phenomenon involving many dynamical variables in extreme parameter ranges. Substantial simplifications are therefore necessary to obtain a suitable statistical sample of magnetic-polarity reversals across a wide range of parameters. In fact, the possibility of reaching extreme parameter regimes provided by shell models has the drawback of losing a detailed description of the field geometry, which should instead be taken into account when interpreting the different magnetic topologies observed in late M dwarfs.  
We sacrificed an accurate description of the magnetic field topology and detailed magnetic field feedback in favor of a nonlinear dynamical description of the system, enabling numerous simulations across a wide range of fundamental parameter values. This trade-off enables us to reach parameter regimes relevant to naturally occurring dynamos without incurring prohibitive computational costs.

Our study primarily focused on the influence of thermal convection on magnetic field variability. However, the reciprocal influence of the magnetic field on convection, particularly when the field intensity becomes significant, has been extensively investigated in the literature. For instance, studies on solar magneto-convection \citep{Fan_Fang_2014ApJ, Karak_et_al_2015A&A, Passos_et_al_2017A&A} have shown that magnetic fields can significantly alter convective flows and differential rotation. Similarly, \citet{Cossette_et_al_2017ApJ} highlighted how magnetic tension can modulate heat transport in solar convection models. Although the mentioned models primarily focus on the action of the magnetic field on convection in $\alpha$-$\Omega$ dynamos, the inverse influence of convection on magnetic fields is not excluded in nature, as thermal and magnetic effects are generally coupled ingredients in the same nonlinear dynamics. The need to investigate this inverse influence is particularly important in the case of $\alpha^2$ dynamos, because these dynamos exhibit magnetoconvective structures that deviate from those typically studied and especially because this effect plays a critical role in the plasma dynamics of the convective envelope.

The model yields abrupt magnetic reversals that spontaneously emerge from the nonlinear coupling between the large-scale magnetic field and the small-scale turbulent modes, particularly near the pitchfork bifurcation threshold. By tuning control parameters such as the Rayleigh and Reynolds numbers, the model naturally transits between different regimes.
The statistical analysis of the magnetic reversals obtained with our simplified $\alpha^2$ dynamo model shows good agreement with the geodynamo polarity reversals. Since the geodynamo is the unique $\alpha^2$ natural dynamo with a data set of polarity reversals suitable for statistical analysis, this agreement shows the reliability of our model in capturing the fundamental dynamics of the $\alpha^2$ dynamo mechanisms, at least from the statistical point of view. However, some limitations remain: magnetic topology features are neglected and can play a key role in specific cases, while density stratification is overlooked and can substantially affect stellar convection \citep{Nishikawa_Kusano_2002ApJ}.
Nevertheless, we note that neglect of density stratification does not make the present results incompatible with those of global models. In particular, \citet{Gastine_Wicht_AA_2012} showed that increasing density stratification can promote multipolar field topologies and enhance temporal variability in convective dynamos. Our approach, which isolates the role of convective heat flux, is therefore complementary rather than contradictory. Both effects, namely density stratification and convective efficiency, are likely to act together in real stars, jointly influencing the dynamo regime and magnetic variability. Moreover, rotation is implicitly accounted for in our model through a flow with maximal kinetic helicity, corresponding to the limit of very low Rossby number (i.e., rapid rotation), typical of fully convective M dwarfs. A more realistic configuration would naturally include both effects, and stronger stratification could further enhance the temporal variability for a given convective efficiency. Future extensions of this work may explore how the convective heat flux behaves in models similar to those of \citet{Gastine_Wicht_AA_2012} when varying the fundamental physical parameters.

The Coriolis force is known to influence the geometry of large-scale convective flows, as shown in global MHD simulations \citep{Brown_et_al_ApJ_2008}. However, in the present reduced magnetoconvection model --- which focuses on the nonlinear coupling between scales and does not retain geometric information --- this term would mainly act as a linear rescaling of the flow amplitude, comparable to increasing the effective thermal driving, and would not modify the fundamental nonlinear behavior or the relative trends among the dynamical regimes explored here. 

One of the significant findings of our model is the positive correlation between the Nusselt number and the frequency of magnetic polarity reversals (Figs.~\ref{fig:vs_Nu} and \ref{fig:number_vs_Nu}). This result highlights the critical role of global convective heat transport in driving magnetic reversals. This result is aligned with geodynamo simulations and paleomagnetic data analysis, where the increase in core-mantle boundary heat flow is considered to be associated with periods of elevated reversal frequency \citep{McFadden_Merrill_1984JGR, Glatzmaier_et_al_1999, Valet_et_al_2005, Lay_et_al_2008, Olson_et_al_2010, Driscoll_Olson_2011, Biggin_et_al_2012, Olson_Hagay_2015, Nakagawa_2020PEPS, Dannberg_et_al_2024GeoJI}. 

However, it is important to clarify that an analogy between Earth’s and the fully convective star’ dynamo is limited to specific aspects, such as the role of heat flux variability in driving magnetic reversals and the $\alpha^2$ dynamo mechanism under certain conditions.
In the spirit of this similarity \citep{Christensen_et_al_2009Natur}, the results of our model can be extended to fully convective stars, whereby significant time variability might be associated with high levels of convective heat flux within their convection zones \citep{Nigro_2022}. 
Observations of rapidly rotating fully convective stars with similar masses and spectral types (e.g., late M dwarfs) have revealed two distinct magnetic regimes \citep{Morin_et_al_2011MNRAS}: a more active branch characterized by high levels of time variability and a branch of a more stable dipolar dynamo \citep{Shulyak_2017NatAs}. 
This dichotomy has been interpreted as resulting from dynamo bistability, whereby the initial condition only determines and locks fully convective stars with similar stellar parameters into one of the two magnetic regimes \citep{Gastine_et_al_2013}. 
The idea of dynamo bistability requires further validation through alternative models and simulations to establish its theoretical foundation. However, the hypothetical bistability dynamo may remain a mere speculative idea if a robust justification for different initial conditions during the early stages of this type of star is not provided.
The idea of dynamo bistability is not accepted by the entire community. Indeed, for instance, \citet{Raedler_etal_1990AA, Moss_Sokoloff_2009AA} adopting numerical models of the mean-field dynamo with an isotropic $\alpha$ effect, obtained results that turn out to be incompatible with the idea of dynamo bistability. Also, more recently, \citet{Kitchatinov_et_al_2014MNRAS} argued that two stable dynamo regimes are improbable in the same parameter range. Adopting a mean-field dynamo model, they attributed the magnetic dichotomy to two different dynamical states between which fully convective stars can cyclically switch.
In addition, high-resolution global MHD simulations, such as those conducted by \citet{Yadav_et_al_2015ApJ} for an M dwarf, although with a 20-day rotation period, result in a stationary dipole-like global field configuration with an average total field strength smaller than that observed. 
Since these simulations yielded only one type of stable solution --- a steady dipolar configuration --- they do not support the hypothesis of a bistable dynamo as an explanation for the observed magnetic dichotomy in late-type M dwarfs. 

Finally, we stress that our results provide a qualitative investigation of $\alpha^2$ dynamo variability under simplified physical assumptions. Despite its limitations, the model offers valuable insight into how convective heat transport influences magnetic variability in natural dynamos, providing a complementary perspective to more complex numerical simulations. 
We propose a possible explanation for the observed magnetic dichotomy without invoking dynamo bistability. Based on our numerical results, we suggest that the key ingredient determining the magnetic regime is the heat flow efficiency \citep{Nigro_2022}: higher heat fluxes lead to more frequent magnetic reversals, whereas lower fluxes favor more stable dipolar states. Therefore, higher magnetic variability characterizes systems with stronger convective heat fluxes. 
However, the heat flux conditions of a given late M star may vary over time, leading to transitions between different magnetic regimes. Changes in the heat flux can, in principle, occur due to variations in plasma microphysics, which consequently lead to changes in the dissipative effects. Moreover, we do not exclude the possibility of modifications in the boundary conditions of the star's convection zone and the surrounding environment of M dwarfs, which could also lead to variations in the heat flux, although these aspects were not explicitly modeled in this study.
Therefore, changes in the plasma condition of the late M star's convective zone can produce dynamic transitions between the two regimes of the magnetic dichotomy over time. This latter scenario is plausible for a late M dwarf, and our description contemplates it by accounting for values of dissipative coefficients that vary over time, offering a flexible and physically motivated alternative to the bistability hypothesis. 

Our results appear to be aligned with magnetic activity variations observed by \citet{Irving_Saar_et_al_2023ApJ} in some M-type stars and by \citet{Route_2016ApJ} in ultra-cool dwarfs, specifically those of spectral type $\geq$ M7 and brown dwarfs. In particular, \citet{Irving_Saar_et_al_2023ApJ} found evidence of a decrease in the characteristic time scale of activity variation for M--type stars with shorter convective turnover times, indicative of a more efficient convective heat flux. Interestingly, they also observed a similar trend in faster-rotating partly convective stars, suggesting that a thermally driven dynamo can partially influence early M-type and faster-rotating FGK--type stars.

Fully convective stars, being among the fastest rotators, experience strong Coriolis forces that enhance helical turbulence and kinetic helicity, likely reaching their highest values among M-type stars. Our model consistently has a notably high level of kinetic helicity. This circumstantial evidence opens the way for future exploration of the relationship between rotational and activity periods in slower-rotating M dwarfs, considering lower kinetic helicity values.

In summary, while the bistability hypothesis attributes the magnetic dichotomy to static initial conditions, effectively excluding the possibility for an M dwarf to switch between different regimes, our study emphasizes the dynamic role of convective heat flux variations. 
This approach overcomes key limitations of the bistability framework by accounting for the intrinsic thermal variability of stellar interiors, which can naturally drive transitions between magnetic regimes as their thermodynamic properties evolve. 

Additional models and DNS of fully convective stars, along with observational studies on M dwarfs, will be essential to validate and refine our findings on the positive correlation between heat flux and magnetic reversal frequency. While our approach is well-suited for exploring extreme parameter regimes and useful for understanding the temporal variability of magnetic fields, DNS are indispensable for investigating the detailed magnetic topologies associated with the two branches of the magnetic dichotomy. A complementary use of different methods might provide a comprehensive understanding of the stellar dynamo mechanisms, bridging the gap between simplified models and DNS.
Future observational campaigns and advanced numerical modeling will hence be essential to understand stellar magnetic variability and to elucidate differences and similarities between their magnetism and solar-like star magnetism \citep{Wright_Drake_2016Natur} from the perspective of a general picture of the astrophysical dynamo mechanism.

\begin{acknowledgments}
We are very grateful to the anonymous reviewer who offered constructive comments and  suggestions that improved the article. 
This work is supported by the MELODY research project, funded by the Italian Ministry of Universities and Research (MUR) under the PNRR Young Researchers program 2022 (MELODY SoE project, grant agreement No SOE\_0000119, CUP E53C22002450006).
\end{acknowledgments}

\appendix
\section{Appendix}
\label{sec:ApexA}
The magneto-convection for an incompressible fluid plasma, i.e. a magnetized fluid with divergence-free velocity $\nabla \cdot {\mathbf{u}} = 0$, can be described by the set of equations governing hydromagnetic Boussinesq convection in a layer of thickness $L$ along the $z$ direction, confined between two horizontal layers at sufficiently large temperature difference $\Delta T$. These equations can be written as follows \citep{Chandrasekhar_book}:

\begin{align}
&\left( \frac{\partial}{\partial t} - \nu \nabla^2 \right) {\bf u} = - \nabla P_{tot} - {\bf u} \cdot \nabla {\bf u} + {\bf b} \cdot \nabla {\bf b} - 
\hat{\alpha} \; \Theta \; {\bf g}
\label{velocity}
\\
&\left( \frac{\partial}{\partial t} - \eta \nabla^2 \right) {\bf b} = - {\bf u} \cdot \nabla {\bf b} + {\bf b} \cdot \nabla {\bf u} 
\label{mag_field}
\\
&\left( \frac{\partial}{\partial t} - \chi \nabla^2 \right){\Theta} = - {\bf u} \cdot \nabla \Theta 
\label{theta}
\end{align}

where ${\bf b}$ is the magnetic field measured in units of the equivalent Alfv\'en speed, i.e. ${\bf B} / \sqrt{4 \pi  \rho}$ being $\rho$ the mass density, $\Theta = T -  \bar {T}  $ is the temperature field fluctuation in the layer diminished by the average temperature of the unperturbed system (i.e., in the absence of any forcing, namely ${\bf u} = 0 $), 
$P_{tot}$ is the total pressure (the sum of kinetic and magnetic pressure), $\hat{\alpha}$ is the volume expansion coefficient, ${\bf g} = (0,0, -g)$ is the gravitational acceleration, $\nu$, $\eta$ and $\chi$ are kinematic viscosity, magnetic diffusivity, and thermal diffusivity, respectively. It is customary to refer to $\bar {T}  $ as the ambient mean value or the linear background temperature profile. Usually, $\bar {T} =  \bar {T}(z) $ such as $d\bar {T}/dz$ is the adverse temperature gradient imposed from outside \citep{Chandrasekhar_book, Verma_book}. 

The Boussinesq approximation, adopted here, assumes that density variations are negligible. This simplification reduces computational complexity and focuses on the key dynamical features of the system. However, it limits the direct applicability of our results to systems in which density stratification does not significantly affect convection.

The above equations (\ref{velocity})-(\ref{mag_field})-(\ref{theta}) can be numerically solved by introducing dimensionless variables as indicated in Section \ref{results}.

In the horizontal directions, we assume periodic boundary conditions. Along the vertical direction, we consider idealized boundary conditions such that the temperature fluctuations, the normal component of the velocity, the tangential component of the stress tensor, and the horizontal component of the magnetic field vanish on the upper and lower boundaries \citep{Cattaneo_et_al_2003ApJ}. These conditions are consistent with periodic boundary conditions \citep{Kerr_periodicBC_1996} such that the original physical equations (\ref{velocity})--(\ref{mag_field})--(\ref{theta}) can be written in the Fourier vector space. 
As highlighted in many papers (see, for instance, \citet{Cattaneo_et_al_2003ApJ}), in the incompressible case, pressure plays the role of keeping the velocity field solenoidal. 
The pressure, in turn, can be formally eliminated from the equations by requiring that the solution be a solenoidal function or that it be orthogonal in the phase space to the wave vector $\bf{k}$ for which the equations are considered. 
In Fourier space, the nonlinear terms take the shape of convolution terms that couple each Fourier mode with all the others. An accurate description of a fluid system with high Reynolds numbers requires many Fourier modes, which is often not feasible with current computational resources.
Shell models allow us to circumvent the latter difficulty. Notably, to reduce the number of modes involved in the nonlinearities, keeping their most peculiar features, a sort of sampling is performed in Fourier space, bearing in mind that the nonlinearities have to preserve the conservation of the right quadratic invariants as they do in the original MHD equations.
Specifically, the shell model equations have been derived considering in the Fourier space $N$ the exponential spacing of shells in $k$. Each shell $k_n = k_0 2^n$ (where $n=1,...,N$ and $k_0 = \pi/L $) encompasses the wave numbers with modulus $k$ such that $k_0 2^{n} < k < k_0 2^{n+1}$. 
The velocity, the magnetic field fluctuations, and the temperature increments on scale $1/k_n$ are represented by complex scalars $u_n$, $b_n$, and $\theta_n$, respectively. We adopt complex scalars to describe phase transfer in turbulent dynamics. We adopt the Boussinesq approximation. Therefore, the MHD shell model equations that we consider are (\ref{eq:GOY_u})-(\ref{eq:GOY_b})-(\ref{eq:Thermal}).
The latter equations represent the extension of the model by \citet{Mingshun_Shida_1997}, originally developed for thermal convection, to which we add the equation for the magnetic field fluctuations to account for the magnetic component of the phenomenon (see also \citep{Brandenburg_1992} and \citep{Jensen_et_al_1992}). 
Considering exponential spacing in $k$, a wide range of scales can be covered, and thus very high Reynolds and Rayleigh numbers can be reached. The shell model equations are obtained by assuming local interaction among scalar fields on the nearest and next-nearest shell neighbors.
Due to the assumption of local interaction between shells, the classical shell model \citep[e.g.][]{Gledzer_1973, Ohkitani_Yamada_1989} is not able to describe the nonlocal action of small-scale turbulence on the large-scale magnetic field, which is the crucial ingredient of mean-field electrodynamics in the scale separation approach \citep{Parker_1955, steenbeck_1966, Moffatt_1978, Krause_Radler_1980}.
Our formulation aims to align with the mean-field electrodynamics theory, which relies on the concept of scale separation. To bridge this gap, we have modified the evolution equation of the large-scale magnetic field (i.e., the magnetic field on the first shell) to incorporate the non-local effect of the small-scale field fluctuations. 

In general, the mean-field electrodynamic theory (MFE) describes the nonlocal effect of small scales on the large-scale magnetic field through the electromotive force $\bf{\epsilon}$ (e.m.f.) acting in the equation of the large-scale magnetic field vector. The e.m.f. can be written as a series expansion of the large-scale magnetic field vector $\langle{\bf b} \rangle$ where the first term is given by
\begin{equation}
\label{eq:alpha_A_ij}
{\epsilon}_i= \alpha_{i, j} \; \langle b \rangle_j \ ,
\end{equation}
for each component of the field ($i = x, y, z$) and, in general, being $\alpha_{i, j}$ a pseudo-tensor whose elements depend on the small-scale properties of the flow. This term is known to produce the $\alpha$ effect. In homogeneous and weakly anisotropic turbulence, the pseudotensor $\alpha_{i, j}$ reduces to a simple scalar (i.e. $\alpha_{i,j} = \delta_{i,j} \; \alpha$ ), and equation \ref{eq:alpha_A_ij} becomes  
\begin{equation}
\label{eq:alpha_A_2}
{\epsilon}_i= \alpha \; \langle b \rangle_i \; ,
\end{equation} 
for each component $i$ of the vector $\langle{\bf b}\rangle$. In the specific case where we adopt a shell model, the large-scale magnetic field $b_1$ is a complex scalar and equation \ref{eq:alpha_A_2} turns in
\begin{equation}
\label{eq:alpha_A}
{\epsilon}= \alpha \; b_1 \; .
\end{equation} 

In the quasi-linear approximation of the MFE, the series expansion of the e.m.f is truncated at the first term, and a parameterization of $\alpha$ is generally adopted to achieve closure and reproduce observations. In particular, solar mean-field dynamo models aim to reproduce solar observations by providing a prescribed $\alpha$ shape and adding a specific $\Omega$ effect. 
In mean field dynamo models, the $\alpha$ parameter is a source term in the large-scale magnetic equation, and, in the linear approximation (i.e., $\alpha$ is a constant parameter), it produces an exponential growth of the large-scale magnetic field ${b_1}$.

In nature, when the magnetic field grows and its amplitude approaches sufficiently large values, the system self-organizes to contrast this growth, avoiding infinite magnetic field amplitude. Indeed, the Lorentz force back-reacts against the same source that generated it. The system thus sets into a saturated magnetic energy regime. 
Therefore, non-linear back-reaction, known as $\alpha$-quenching, is generally adopted in large-scale dynamo systems. 

The expression of $\alpha$-quenching, and thus its saturation level, is under debate (see for more details \citet{Tobias_2021JFM} and references therein). In our model, we assume that the energy equipartition is achieved in the saturation regime, which is consistent with different dynamo models \citep[e.g.,][etc.]{Biermann_et_al_1951PhRv, Charbonneau_MacGregor_1996ApJ, Kulsrud_et_al_1997ApJ, Brandenburg_Subramanian_2000A&A, Archontis_et_al_2007AA, Cho_et_al_2009ApJ, Beresnyak_2012PhRvL} and turbulent dynamo experiments \citep{Tzeferacos_2018NatCo}.
We assume an $\alpha$ quenching that works efficiently when the magnetic energy (per unit volume) is comparable to the kinetic energy of fluid motions, that is, when $B_0 \approx \langle |{\bf u}|^2 \rangle $, and $B_0$ is usually denoted as the equipartition field strength.
Moreover, we note that the $\alpha$ quenching has to depend on the strength of the mean magnetic field ${b_1}$ because the back-reaction of the mean field acts by contrasting its own growth and becomes stronger for increasing magnetic field strength (for more details on the mean-field dynamo theory and $\alpha$-quenching, see \citet{Biskamp_book, Brandenburg_Subramanian_2005PhR, Charbonneau_2010LRSP, Charbonneau_2020LRSP}). 
Since the e.m.f. in the induction equation must cause the magnetic field to change sign, the $\alpha \; {b_1}$ term in the equation (\ref{eq:alpha_A}) must be an odd function of ${b_1}$ \citep{Field_Blackman_2002ApJ}. This means that the $\alpha$ parameter in Equation (\ref{eq:alpha_A}) had to depend on $b_1^2$. 
In particular, considering the second-order correlation tensor for the small-scale field fluctuations \citep{Rudiger_1973, Rudiger_1974, Roberts_Soward_1975, Moffatt_72}, we can consider the following $\alpha$-quenching:
\begin{equation}
\label{eq:alpha1_A}
\alpha=\mu \left(1- \frac{{b}_1^2}{B^2_{0}}\right) \ ,
\end{equation}
where $B_0$ is the amplitude of the equipartition or saturation magnetic field, and $\mu$ is a constant related to the kinetic helicity of the flow motions, which is at its maximum value defined by the construction of the shell model. We therefore set $\mu = 1$ considering the typical value of normalized kinetic helicity $H_k \approx 1$ as discussed in Section \ref{Model}.

To summarize, we adopt the scale separation approach; we consider the quasilinear approximation of the $\alpha$ term on the large-scale magnetic field equation (i.e. the equation of the magnetic field on the first shell), where the turbulent electromotive force is approximated by equation (\ref{eq:alpha_A}), and $\alpha$ is provided by equation (\ref{eq:alpha1_A}). Therefore, the resulting equation for the magnetic field in the first shell is given by equation (\ref{eq:b1}), which has to be solved together with the other shell model equations (\ref{eq:GOY_u})--(\ref{eq:GOY_b})--(\ref{eq:Thermal}).

%
%
\section{Appendix}
\label{sec:Apexb}
\begin{figure*}
\centering 
\subfloat[The frequency of the magnetic field $b_1(t)$ polarity reversals increases with increasing Rayleigh number (increasing $\tilde{\alpha}$), keeping constant the other parameters.]{%
  \includegraphics[width=0.99\textwidth]{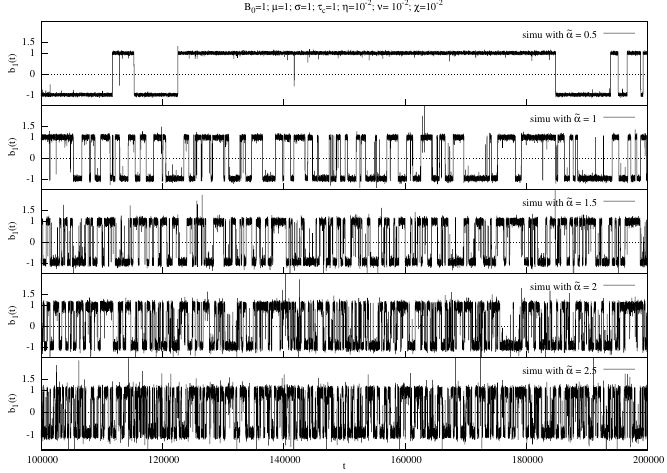}%
  \label{fig:vs_alpha_dyn}}\qquad
\subfloat[The PDFs on the left show the tendency to develop shorter persistence times for an increasing $\tilde{\alpha}$. On the right, the sign singularity in the magnetic reversals is revealed by Partition Functions $\zeta$.  
 On the left, a linear fit is made in the PDF wings (neglecting the persistence times longer than 20000 unit time) with the slope indicated in the corresponding area of the fitting line with its corresponding error. 
On the right, the linear fit is made in a suitable inertial range for each partition function depicted with slopes and errors, respectively, indicated in the panel. 
For comparison, we also show the partition function computed from the CK95 dataset with corresponding slope and error.]{%
  \includegraphics[width=0.99 \textwidth]{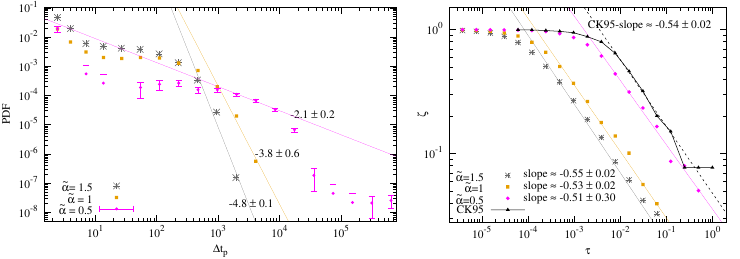}%
  \label{fig:vs_alpha}%
}
\caption{Sensitivity of the system with respect to the thermal convection coefficient of the velocity $\tilde{\alpha}$ (i.e., Rayleigh number $\textrm{Ra} = {\tilde{\alpha}}{\theta_0}L^3/(\nu \chi))$), keeping constant the other parameters as indicated in the figure title.}
\label{fig:whole_vs_alpha}
\end{figure*}

\begin{figure*}
\centering 
\subfloat[The frequency of the magnetic field $b_1(t)$ polarity reversals increases with increasing Reynolds number.]{%
  \includegraphics[width=0.99\textwidth]{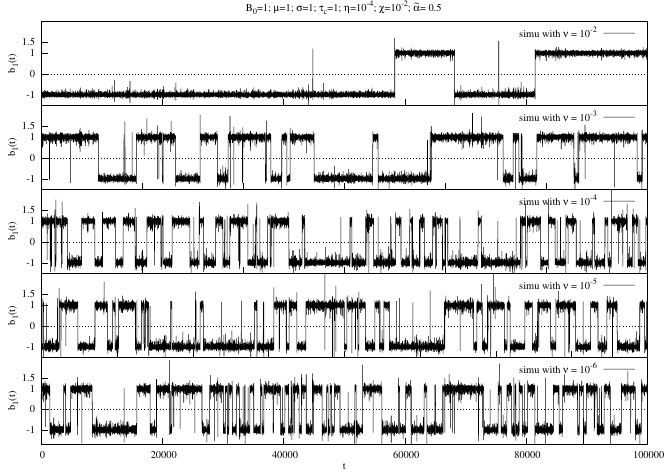}%
  \label{fig:vs_nu_dyn}%
}\qquad
\subfloat[Tendency to develop shorter persistence times for an increasing Reynolds number Re, keeping constant the other parameters,
     shown by the PDFs on the left, and sign-singularity in the magnetic reversals as revealed by Partition Functions $\zeta$, on the right.  
  On the left, a linear fit is made in the PDF wings (neglecting the persistence times longer than 20000 unit time) with the slope indicated in the corresponding area of the fitting line with its corresponding error. 
 On the right, the linear fit is made in a suitable inertial range for each partition function depicted with slopes and errors, respectively, indicated in the panel. 
We also computed the partition function on the CK95 dataset with slope and error, as shown in the plot.]{%
  \includegraphics[width=0.99\textwidth]{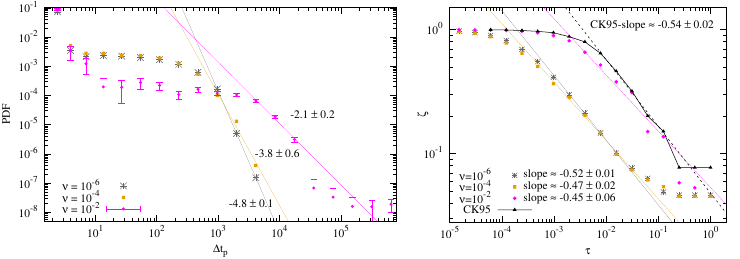}%
  \label{fig:vs_nu}%
}
\caption{Sensitivity of the system with respect to the kinematic viscosity $\nu$ (i.e., Reynolds number Re$=Lu_0/\nu$), keeping constant the other parameters; in particular, $\tilde{\alpha}=0.5$, $\chi = 10^{-2}$, and $\eta = 10^{-4}$.}
\label{fig:whole_vs_nu}
\end{figure*}

\begin{figure*}
\centering 
\subfloat[Considering the same interval of time, the number of the magnetic field $b_1(t)$ polarity reversals increases for increasing magnetic Reynolds number.]{%
  \includegraphics[width=0.99\textwidth]{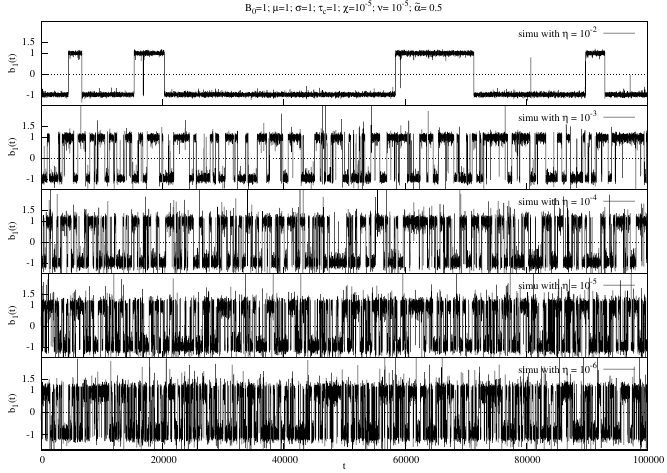}%
  \label{fig:vs_eta_dyn}%
}\qquad
\subfloat[The tendency to develop shorter persistence times for an increasing magnetic Reynolds number is shown by the  
    PDFs on the left, and the sign-singularity in the magnetic reversals, as revealed by Partition Functions $\zeta$, on the right.  
 On the left, a linear fit is made in the PDF wings with the slope indicated in the corresponding area of the fitting line with its corresponding error. Black triangles represent the PDF inferred from the CK95 database for the Earth's polarity reversals, which are being introduced for comparison.  
On the right, the linear fit is made in a suitable inertial range for each partition function depicted with slopes 
$-0.49 \pm 0.02$, $-0.50 \pm 0.02$, and  $-0.52 \pm 0.01$ when $\eta = 10^{-2}$, $\eta = 10^{-4}$ and $\eta = 10^{-6}$, 
respectively, in the simulations obtained by setting $\chi = 10^{-5}$ and $\nu = 10^{-5}$.  
We also computed the partition function on the CK95 dataset with slope and error, as shown in the plot.]{%
  \includegraphics[width=0.99\textwidth]{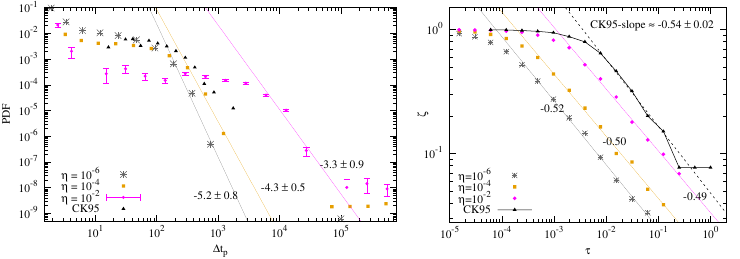}%
  \label{fig:vs_eta}%
}
\caption{Sensitivity of the system with respect to the $\eta$ diffusivity (i.e., magnetic Reynolds number Rm$=Lu_0/\eta$), keeping constant the other parameters.}
\label{fig:whole_vs_eta}
\end{figure*}

In this section, we provide additional simulations showing the characteristic behavior of the large-scale magnetic field that alternates its sign, thus producing magnetic polarity reversals (see Fig.~\ref{fig:vs_alpha_dyn}, Fig.~\ref{fig:vs_nu_dyn} and Fig.~\ref{fig:vs_eta_dyn}. The persistence times are thus statistically analyzed (see Fig.~\ref{fig:vs_alpha}, Fig.~\ref{fig:vs_nu} and Fig.~\ref{fig:vs_eta}). The simulations depicted here correspond to the models considered in Table \ref{tab:slopes}

%
%

\bibliography{biblio.bib}{}

@BOOK{Chandrasekhar_book,
       author = {{Chandrasekhar}, Subrahmanyan},
        title = "{Hydrodynamic and hydromagnetic stability}",
         year = 1961,
       adsurl = {https://ui.adsabs.harvard.edu/abs/1961hhs..book.....C}
}

@book{Verma_book,
    author={Verma, K.M.},
    title ={Physics of Buoyant Flows},
    series = {World Scientific},
    year = {2018},
    publisher = {World Scientific Publishing Co. Pte. Ltd.},
    url  = {arXiv: https://www.worldscientific.com/doi/pdf/10.1142/10928}
}

@book{White_1988,
   author = {White, F.M.}, 
   title = {Heat and Mass Transfer},
    series = {Addison-Wesley series in mechanical engineering},
    year = {1988},
    publisher = {Reading, Mass.: Addison-Wesley (MA, USA)},
}

@book{Ozisik_1985,
   author = {\"{O}zisik, M.N. }, 
   title = {Heat Transfer},
    series = {},
    year = {1985},
    publisher = {McGraw-Hill International Editions (New York, NY, USA)},
}

@BOOK{Krause_Radler_1980,
       author = {{Krause}, F. and {Raedler}, K. -H.},
        title = "{Mean-field magnetohydrodynamics and dynamo theory}",
         year = 1980,
       adsurl = {https://ui.adsabs.harvard.edu/abs/1980opp..bookR....K}
}

@BOOK{Merrill_et_al,
       author = {{Merrill}, Ronald T. and {McElhinny}, Michael W. and {McFadden}, Phillip L},
        title = "{The Magnetic Field of the Earth: Paleomagnetism, the Core, and the Deep Mantle}",
      publisher = {Academic Press, San Diego, Calif.},
         year = {1996},
}

@BOOK{Milankovictch_1969ciip_book,
       author = {{Milankovictch}, Milutin},
        title = "{Canon of insolation and the ice-age problem (Kanon der Erdbestrahlung und seine Anwendung auf das Eiszeitenproblem) Belgrade, 1941.}",
         year = 1969,
       adsurl = {https://ui.adsabs.harvard.edu/abs/1969ciip.book.....M}
}

@ARTICLE{Gloaguen_et_al_1985,
       author = {{Gloaguen}, C. and {L{\'e}orat}, J. and {Pouquet}, A. and {Grappin}, R.},
        title = "{A scalar model for MHD turbulence}",
      journal = {Physica D Nonlinear Phenomena},
         year = 1985,
        month = oct,
       volume = {17},
       number = {2},
        pages = {154-182},
          doi = {10.1016/0167-2789(85)90002-8},
       adsurl = {https://ui.adsabs.harvard.edu/abs/1985PhyD...17..154G}
}

@ARTICLE{Plunian_et_al_2013,
       author = {{Plunian}, Franck and {Stepanov}, Rodion and {Frick}, Peter},
        title = "{Shell models of magnetohydrodynamic turbulence}",
      journal = {\physrep},
         year = 2013,
        month = feb,
       volume = {523},
       number = {1},
        pages = {1-60},
          doi = {10.1016/j.physrep.2012.09.001},
archivePrefix = {arXiv},
       eprint = {1209.3844},
 primaryClass = {physics.flu-dyn},
       adsurl = {https://ui.adsabs.harvard.edu/abs/2013PhR...523....1P}
}

@ARTICLE{Biferale_2003,
       author = {{Biferale}, Luca},
        title = "{Shell Models of Energy Cascade in Turbulence}",
      journal = {Annual Review of Fluid Mechanics},
         year = 2003,
        month = jan,
       volume = {35},
       number = {35},
        pages = {441-468},
          doi = {10.1146/annurev.fluid.35.101101.161122},
       adsurl = {https://ui.adsabs.harvard.edu/abs/2003AnRFM..35..441B}
}

@ARTICLE{Nigro_2013GApFD,
       author = {{Nigro}, Giuseppina},
        title = "{A shell model for a large-scale turbulent dynamo}",
      journal = {Geophysical and Astrophysical Fluid Dynamics},
         year = 2013,
        month = feb,
       volume = {107},
       number = {1-2},
        pages = {101-113},
          doi = {10.1080/03091929.2012.664141},
       adsurl = {https://ui.adsabs.harvard.edu/abs/2013GApFD.107..101N}
}

@ARTICLE{Nigro_2011ApJ,
       author = {{Nigro}, Giuseppina and {Veltri}, Pierluigi},
        title = "{A Study of the Dynamo Transition in a Self-consistent Nonlinear Dynamo Model}",
      journal = {\apjl},
         year = 2011,
        month = oct,
       volume = {740},
       number = {2},
          eid = {L37},
        pages = {L37},
          doi = {10.1088/2041-8205/740/2/L37},
archivePrefix = {arXiv},
       eprint = {1010.5213},
 primaryClass = {astro-ph.EP},
       adsurl = {https://ui.adsabs.harvard.edu/abs/2011ApJ...740L..37N}
}

@ARTICLE{Suzuki_Sadayoshi_1995,
       author = {{Suzuki}, Eri and {Toh}, Sadayoshi},
        title = "{Entropy cascade and temporal intermittency in a shell model for convective turbulence}",
      journal = {\pre},
         year = 1995,
        month = jun,
       volume = {51},
       number = {6},
        pages = {5628-5635},
          doi = {10.1103/PhysRevE.51.5628},
       adsurl = {https://ui.adsabs.harvard.edu/abs/1995PhRvE..51.5628S}
}

@ARTICLE{Gledzer_1973,
       author = {{Gledzer}, E.~B.},
        title = "{System of hydrodynamic type admitting two quadratic integrals of motion}",
      journal = {Soviet Physics Doklady},
         year = 1973,
        month = oct,
       volume = {18},
        pages = {216},
       adsurl = {https://ui.adsabs.harvard.edu/abs/1973SPhD...18..216G}
}

@ARTICLE{Olson_Hagay_2015,
       author = {{Olson}, Peter and {Hagay}, Amit},
        title = "{Mantle superplumes induce geomagnetic superchrons}",
      journal = {Frontiers in Earth Science},
         year = 2015,
        month = jul,
       volume = {3},
          eid = {38},
        pages = {38},
          doi = {10.3389/feart.2015.00038},
       adsurl = {https://ui.adsabs.harvard.edu/abs/2015FrEaS...3...38O}
}

@ARTICLE{Ohkitani_Yamada_1989,
       author = {{Ohkitani}, K. and {Yamada}, M.},
        title = "{Temporal Intermittency in the Energy Cascade Process and Local Lyapunov Analysis in Fully-Developed Model Turbulence}",
      journal = {Progress of Theoretical Physics},
         year = 1989,
        month = feb,
       volume = {81},
       number = {2},
        pages = {329-341},
          doi = {10.1143/PTP.81.329},
       adsurl = {https://ui.adsabs.harvard.edu/abs/1989PThPh..81..329O}
}

@ARTICLE{Dannberg_et_al_2024GeoJI,
       author = {{Dannberg}, Juliane and {Gassm{\"o}ller}, Rene and {Thallner}, Daniele and {LaCombe}, Frederick and {Sprain}, Courtney},
        title = "{Changes in core-mantle boundary heat flux patterns throughout the supercontinent cycle}",
      journal = {Geophysical Journal International},
         year = 2024,
        month = jun,
       volume = {237},
       number = {3},
        pages = {1251-1274},
          doi = {10.1093/gji/ggae075},
archivePrefix = {arXiv},
       eprint = {2310.03229},
 primaryClass = {physics.geo-ph},
       adsurl = {https://ui.adsabs.harvard.edu/abs/2024GeoJI.237.1251D}
}

@ARTICLE{Nakagawa_2020PEPS,
       author = {{Nakagawa}, Takashi},
        title = "{A coupled core-mantle evolution: review and future prospects}",
      journal = {Progress in Earth and Planetary Science},
         year = 2020,
        month = dec,
       volume = {7},
       number = {1},
          eid = {57},
        pages = {57},
          doi = {10.1186/s40645-020-00374-8},
       adsurl = {https://ui.adsabs.harvard.edu/abs/2020PEPS....7...57N}
}

@ARTICLE{Charbonneau_2010LRSP,
       author = {{Charbonneau}, Paul},
        title = "{Dynamo Models of the Solar Cycle}",
      journal = {Living Reviews in Solar Physics},
         year = 2010,
        month = sep,
       volume = {7},
       number = {1},
          eid = {3},
        pages = {3},
          doi = {10.12942/lrsp-2010-3},
       adsurl = {https://ui.adsabs.harvard.edu/abs/2010LRSP....7....3C}
}

@ARTICLE{Brandenburg_Subramanian_2005PhR,
       author = {{Brandenburg}, Axel and {Subramanian}, Kandaswamy},
        title = "{Astrophysical magnetic fields and nonlinear dynamo theory}",
      journal = {\physrep},
         year = 2005,
        month = oct,
       volume = {417},
       number = {1-4},
        pages = {1-209},
          doi = {10.1016/j.physrep.2005.06.005},
archivePrefix = {arXiv},
       eprint = {astro-ph/0405052},
 primaryClass = {astro-ph},
       adsurl = {https://ui.adsabs.harvard.edu/abs/2005PhR...417....1B}
}

@ARTICLE{Jensen_et_al_1992,
       author = {{Jensen}, M.~H. and {Paladin}, G. and {Vulpiani}, A.},
        title = "{Shell model for turbulent advection of passive-scalar fields}",
      journal = {\pra},
         year = 1992,
        month = may,
       volume = {45},
       number = {10},
        pages = {7214-7221},
          doi = {10.1103/PhysRevA.45.7214},
       adsurl = {https://ui.adsabs.harvard.edu/abs/1992PhRvA..45.7214J}
}

@ARTICLE{Brandenburg_1992,
       author = {{Brandenburg}, Axel},
        title = "{Energy spectra in a model for convective turbulence}",
      journal = {\prl},
         year = 1992,
        month = jul,
       volume = {69},
       number = {4},
        pages = {605-608},
          doi = {10.1103/PhysRevLett.69.605},
       adsurl = {https://ui.adsabs.harvard.edu/abs/1992PhRvL..69..605B}
}

@ARTICLE{Frick_Sokoloff_1998,
       author = {{Frick}, Peter and {Sokoloff}, Dmitriy},
        title = "{Cascade and dynamo action in a shell model of magnetohydrodynamic turbulence}",
      journal = {\pre},
         year = 1998,
        month = apr,
       volume = {57},
       number = {4},
        pages = {4155-4164},
          doi = {10.1103/PhysRevE.57.4155},
       adsurl = {https://ui.adsabs.harvard.edu/abs/1998PhRvE..57.4155F}
}

@BOOK{Moffatt_1978,
       author = {{Moffatt}, H.~K.},
        title = "{Magnetic field generation in electrically conducting fluids}",
         year = 1978,
       adsurl = {https://ui.adsabs.harvard.edu/abs/1978mfge.book.....M}
}

@ARTICLE{Mingshun_Shida_1997,
       author = {{Mingshun}, Jiang and {Shida}, Liu},
        title = "{Scaling behavior of velocity and temperature in a shell model for thermal convective turbulence}",
      journal = {\pre},
         year = 1997,
        month = jul,
       volume = {56},
       number = {1},
        pages = {441-446},
          doi = {10.1103/PhysRevE.56.441},
       adsurl = {https://ui.adsabs.harvard.edu/abs/1997PhRvE..56..441M}
}

@ARTICLE{Parker_1955,
       author = {{Parker}, Eugene N.},
        title = "{Hydromagnetic Dynamo Models.}",
      journal = {\apj},
         year = 1955,
        month = sep,
       volume = {122},
        pages = {293},
          doi = {10.1086/146087},
       adsurl = {https://ui.adsabs.harvard.edu/abs/1955ApJ...122..293P}
}

@ARTICLE{Steenbeck_1966,
       author = {{Steenbeck}, M. and {Krause}, F. and {R{\"a}dler}, K. -H.},
        title = "{Berechnung der mittleren Lorentz-Feldst\"aarke f\"ur ein elektrisch leitendes Medium in turbulenter, durch Coriolis-Kr\"afte beeinflu{\ss}ter Bewegung}",
      journal = {Zeitschrift Naturforschung Teil A},
         year = 1966,
        month = apr,
       volume = {21},
        pages = {369},
          doi = {10.1515/zna-1966-0401},
       adsurl = {https://ui.adsabs.harvard.edu/abs/1966ZNatA..21..369S}
}

@ARTICLE{Rudiger_1973,
       author = {{R{\"u}diger}, G.},
        title = "{Behandlung eines Einfachen Hydromagnetischen Dynamos mittels Lineariserung}",
      journal = {Astronomische Nachrichten},
         year = 1973,
        month = jul,
       volume = {294},
       number = {4},
        pages = {183-186},
          doi = {10.1002/asna.19722940407},
       adsurl = {https://ui.adsabs.harvard.edu/abs/1973AN....294..183R}
}

@ARTICLE{Rudiger_1974,
       author = {{Ruediger}, G.},
        title = "{The influence of a uniform magnetic field of arbitrary strength on turbulence}",
      journal = {Astronomische Nachrichten},
         year = 1974,
        month = jan,
       volume = {295},
       number = {6},
        pages = {275-284},
          doi = {10.1002/asna.19742950605},
       adsurl = {https://ui.adsabs.harvard.edu/abs/1974AN....295..275R}
}

@ARTICLE{Roberts_Soward_1975,
       author = {{Roberts}, P.~H. and {Soward}, A.~M.},
        title = "{A unified approach to mean field electrodynamics}",
      journal = {Astronomische Nachrichten},
         year = 1975,
        month = jan,
       volume = {296},
       number = {2},
        pages = {49-64},
          doi = {10.1002/asna.19752960202},
       adsurl = {https://ui.adsabs.harvard.edu/abs/1975AN....296...49R}
}

@ARTICLE{Moffatt_72,
       author = {{Moffatt}, H.~K.},
        title = "{An approach to a dynamic theory of dynamo action in a rotating conducting fluid}",
      journal = {Journal of Fluid Mechanics},
         year = 1972,
        month = jan,
       volume = {53},
        pages = {385-399},
          doi = {10.1017/S0022112072000205},
       adsurl = {https://ui.adsabs.harvard.edu/abs/1972JFM....53..385M}
}

@BOOK{Biskamp_book,
       author = {{Biskamp}, Dieter},
        title = "{Nonlinear magnetohydrodynamics}",
         year = 1993,
       adsurl = {https://ui.adsabs.harvard.edu/abs/1993noma.book.....B}
}

@ARTICLE{Benzi_2005,
       author = {{Benzi}, Roberto},
        title = "{Flow Reversal in a Simple Dynamical Model of Turbulence}",
      journal = {\prl},
         year = 2005,
        month = jul,
       volume = {95},
       number = {2},
          eid = {024502},
        pages = {024502},
          doi = {10.1103/PhysRevLett.95.024502},
archivePrefix = {arXiv},
       eprint = {nlin/0410048},
 primaryClass = {nlin.CD},
       adsurl = {https://ui.adsabs.harvard.edu/abs/2005PhRvL..95b4502B}
}

@ARTICLE{Cattaneo_et_al_2003ApJ,
       author = {{Cattaneo}, Fausto and {Emonet}, Thierry and {Weiss}, Nigel},
        title = "{On the Interaction between Convection and Magnetic Fields}",
      journal = {\apj},
         year = 2003,
        month = may,
       volume = {588},
       number = {2},
        pages = {1183-1198},
          doi = {10.1086/374313},
       adsurl = {https://ui.adsabs.harvard.edu/abs/2003ApJ...588.1183C}
}

@ARTICLE{Schmalz_Stix_1991,
       author = {{Schmalz}, S. and {Stix}, M.},
        title = "{An alpha-Omega dynamo with order and chaos}",
      journal = {\aap},
         year = 1991,
        month = may,
       volume = {245},
       number = {2},
        pages = {654-661},
       adsurl = {https://ui.adsabs.harvard.edu/abs/1991A&A...245..654S}
}

@ARTICLE{Rohit_Pankaj_2017,
       author = {{Kumar}, Rohit and {Wahi}, Pankaj},
        title = "{Dynamo transition in a five-mode helical model}",
      journal = {Physics of Plasmas},
         year = 2017,
        month = sep,
       volume = {24},
       number = {9},
          eid = {092305},
        pages = {092305},
          doi = {10.1063/1.4998472},
archivePrefix = {arXiv},
       eprint = {1708.06124},
 primaryClass = {physics.flu-dyn},
       adsurl = {https://ui.adsabs.harvard.edu/abs/2017PhPl...24i2305K}
}

@ARTICLE{Yadav_et_al_2010,
       author = {{Yadav}, R. and {Chandra}, M. and {Verma}, M.~K. and {Paul}, S. and {Wahi}, P.},
        title = "{Dynamo transition under Taylor-Green forcing}",
      journal = {EPL (Europhysics Letters)},
         year = 2010,
        month = sep,
       volume = {91},
       number = {6},
        pages = {69001},
          doi = {10.1209/0295-5075/91/69001},
       adsurl = {https://ui.adsabs.harvard.edu/abs/2010EL.....9169001Y}
}

@ARTICLE{Stix_1972,
       author = {{Stix}, M.},
        title = "{Non-Linear Dynamo Waves}",
      journal = {\aap},
         year = 1972,
        month = aug,
       volume = {20},
        pages = {9},
       adsurl = {https://ui.adsabs.harvard.edu/abs/1972A&A....20....9S}
}

@ARTICLE{Meinel_Brandenburg_1990,
       author = {{Meinel}, R. and {Brandenburg}, A.},
        title = "{Behaviour of highly supercritical alpha-effect dynamos}",
      journal = {\aap},
         year = 1990,
        month = nov,
       volume = {238},
       number = {1-2},
        pages = {369-376},
       adsurl = {https://ui.adsabs.harvard.edu/abs/1990A&A...238..369M}
}

@ARTICLE{Tobias_1996,
       author = {{Tobias}, S.~M.},
        title = "{Grand minimia in nonlinear dynamos.}",
      journal = {\aap},
         year = 1996,
        month = mar,
       volume = {307},
        pages = {L21},
       adsurl = {https://ui.adsabs.harvard.edu/abs/1996A&A...307L..21T}
}

@ARTICLE{Tobias_1997,
       author = {{Tobias}, S.~M.},
        title = "{The solar cycle: parity interactions and amplitude modulation.}",
      journal = {\aap},
         year = 1997,
        month = jun,
       volume = {322},
        pages = {1007-1017},
       adsurl = {https://ui.adsabs.harvard.edu/abs/1997A&A...322.1007T}
}

@ARTICLE{Passos_et_al_2012,
       author = {{Passos}, D{\'a}rio and {Charbonneau}, Paul and {Beaudoin}, Patrice},
        title = "{An Exploration of Non-kinematic Effects in Flux Transport Dynamos}",
      journal = {\solphys},
         year = 2012,
        month = jul,
       volume = {279},
       number = {1},
        pages = {1-22},
          doi = {10.1007/s11207-012-9971-2},
       adsurl = {https://ui.adsabs.harvard.edu/abs/2012SoPh..279....1P}
}

@article{Stix_2004,
author={Stix, M.},
title={The Sun: An introduction},
journal={Springer-Verlag},
fjournal={Springer-Verlag},
volume={Berlin},
pages={},
year={2004},
}

@ARTICLE{Nigro_2015,
       author = {{Nigro}, G. and {Carbone}, V.},
        title = "{Finite-time singularities and flow regularization in a hydromagnetic shell model at extreme magnetic Prandtl numbers}",
      journal = {New Journal of Physics},
         year = 2015,
        month = jul,
       volume = {17},
       number = {7},
          eid = {073038},
        pages = {073038},
          doi = {10.1088/1367-2630/17/7/073038},
       adsurl = {https://ui.adsabs.harvard.edu/abs/2015NJPh...17g3038N}
}

@ARTICLE{Nigro_2004,
       author = {{Nigro}, Giuseppina and {Malara}, Francesco and {Carbone}, Vincenzo and {Veltri}, Pierluigi},
        title = "{Nanoflares and MHD Turbulence in Coronal Loops: A Hybrid Shell Model}",
      journal = {\prl},
         year = 2004,
        month = may,
       volume = {92},
       number = {19},
          eid = {194501},
        pages = {194501},
          doi = {10.1103/PhysRevLett.92.194501},
       adsurl = {https://ui.adsabs.harvard.edu/abs/2004PhRvL..92s4501N}
}

@ARTICLE{Cadavid_et_al_2014ApJ,
       author = {{Cadavid}, A.~C. and {Lawrence}, J.~K. and {Christian}, D.~J. and {Jess}, D.~B. and {Nigro}, G.},
        title = "{Heating Mechanisms for Intermittent Loops in Active Region Cores from AIA/SDO EUV Observations}",
      journal = {\apj},
         year = 2014,
        month = nov,
       volume = {795},
       number = {1},
          eid = {48},
        pages = {48},
          doi = {10.1088/0004-637X/795/1/48},
archivePrefix = {arXiv},
       eprint = {1404.7824},
 primaryClass = {astro-ph.SR},
       adsurl = {https://ui.adsabs.harvard.edu/abs/2014ApJ...795...48C}
}

@ARTICLE{Nigro_Carbone_2011,
       author = {{Nigro}, Giuseppina and {Carbone}, Vincenzo},
        title = "{Magnetic reversals in a modified shell model for magnetohydrodynamics turbulence}",
      journal = {\pre},
         year = 2010,
        month = jul,
       volume = {82},
       number = {1},
          eid = {016313},
        pages = {016313},
          doi = {10.1103/PhysRevE.82.016313},
archivePrefix = {arXiv},
       eprint = {1001.1001},
 primaryClass = {astro-ph.SR},
       adsurl = {https://ui.adsabs.harvard.edu/abs/2010PhRvE..82a6313N}
}

@ARTICLE{Dominguez_2020,
       author = {{Dom{\'\i}nguez}, Macarena and {Nigro}, Giuseppina and {Mu{\~n}oz}, V{\'\i}ctor and {Carbone}, Vincenzo and {Riquelme}, Mario},
        title = "{Study of the fractality in a magnetohydrodynamic shell model forced by solar wind fluctuations}",
      journal = {Nonlinear Processes in Geophysics},
         year = 2020,
        month = apr,
       volume = {27},
       number = {2},
        pages = {175-185},
          doi = {10.5194/npg-27-175-2020},
       adsurl = {https://ui.adsabs.harvard.edu/abs/2020NPGeo..27..175D}
}

@ARTICLE{Dominguez_2018,
       author = {{Dom{\'\i}nguez}, Macarena and {Nigro}, Giuseppina and {Mu{\~n}oz}, V{\'\i}ctor and {Carbone}, Vincenzo},
        title = "{Study of the fractality of magnetized plasma using an MHD shell model driven by solar wind data}",
      journal = {Physics of Plasmas},
         year = 2018,
        month = sep,
       volume = {25},
       number = {9},
          eid = {092302},
        pages = {092302},
          doi = {10.1063/1.5034129},
       adsurl = {https://ui.adsabs.harvard.edu/abs/2018PhPl...25i2302D}
}

@ARTICLE{Dominguez_2017,
       author = {{Dom{\'\i}nguez}, M. and {Nigro}, G. and {Mu{\~n}oz}, V. and {Carbone}, V.},
        title = "{Study of fractal features of magnetized plasma through an MHD shell model}",
      journal = {Physics of Plasmas},
         year = 2017,
        month = jul,
       volume = {24},
       number = {7},
          eid = {072308},
        pages = {072308},
          doi = {10.1063/1.4993200},
       adsurl = {https://ui.adsabs.harvard.edu/abs/2017PhPl...24g2308D}
}

@ARTICLE{Munoz_et_al_2018,
       author = {{Mu{\~n}oz}, V{\'\i}ctor and {Dom{\'\i}nguez}, Macarena and {Alejandro Valdivia}, Juan and {Good}, Simon and {Nigro}, Giuseppina and {Carbone}, Vincenzo},
        title = "{Evolution of fractality in space plasmas of interest to geomagnetic activity}",
      journal = {Nonlinear Processes in Geophysics},
         year = 2018,
        month = mar,
       volume = {25},
       number = {1},
        pages = {207-216},
          doi = {10.5194/npg-25-207-2018},
       adsurl = {https://ui.adsabs.harvard.edu/abs/2018NPGeo..25..207M}
}

@article{Halmos_2018,
author={Halmos, P.R.},
title={Measure Theory},
journal={Norstrand},
fjournal={New Yourk},
volume={},
pages={Chap. 6},
year={1962},
}

@article{Ott_et_al_1992,
author={Ott, E. and Du, Y. and Sreenivasan, K.R. and Juneja, A. and Suri,  A.K.},
title={Sign-singular measures: Fast magnetic dynamos, and high-Reynolds-number fluid turbulence},
journal={Physical Review Letters},
fjournal={PRL},
volume={69},
pages={2654-2657},
year={1992},
}

@article{Du_et_al_1994,
author={Du, Y. and T$\acute{e}$l, T. and Ott, E.},
title={Characterization of sign singular measures},
journal={Physica D},
fjournal={Physica D},
volume={76},
pages={168-180},
year={1994},
}

@ARTICLE{Nigro_et_al_2017,
       author = {{Nigro}, G. and {Pongkitiwanichakul}, P. and {Cattaneo}, F. and {Tobias}, S.~M.},
        title = "{What is a large-scale dynamo?}",
      journal = {\mnras},
         year = 2017,
        month = jan,
       volume = {464},
       number = {1},
        pages = {L119-L123},
          doi = {10.1093/mnrasl/slw190},
       adsurl = {https://ui.adsabs.harvard.edu/abs/2017MNRAS.464L.119N}
}

@ARTICLE{Peera_et_al_2016,
       author = {{Pongkitiwanichakul}, P. and {Nigro}, G. and {Cattaneo}, F. and {Tobias}, S.~M.},
        title = "{Shear-driven Dynamo Waves in the Fully Nonlinear Regime}",
      journal = {\apj},
         year = 2016,
        month = jul,
       volume = {825},
       number = {1},
          eid = {23},
        pages = {23},
          doi = {10.3847/0004-637X/825/1/23},
       adsurl = {https://ui.adsabs.harvard.edu/abs/2016ApJ...825...23P}
}

@article{CK95,
author={Cande, S.C. and Kent, D.V.},
title={Revised calibration of the geomagnetic polarity timescale for the Late Cretaceous and Cenozoic},
journal={Journal of Geophysical Research},
fjournal={J. Geophys. Res.},
volume={100},
pages={6093-6095},
year={1995},
}

@article{Sun_et_al_2005,
author={ Sun, C. and Xi, H.-D. and Xia K.-Q.},
title={Azimuthal symmetry, flow dynamics, and heat transport in turbulent thermal convection in a cylinder with an aspect ratio of 0.5},
journal={Physical Review Letters},
fjournal={PRL},
volume={95},
pages={074502},
year={2005},
}

@article{Xi_et_al_2008,
author={ Xi H.-D. and Xia K.-Q.},
title={Flow mode transitions in turbulent thermal convection},
journal={Physics of Fluids},
fjournal={Phys. Fluids},
volume={20},
pages={055104},
year={2008},
}

@ARTICLE{Weiss_Ahlers_2011,
       author = {{Weiss}, Stephan and {Ahlers}, Guenter},
        title = "{Turbulent Rayleigh-B{\'e}nard convection in a cylindrical container with aspect ratio {\ensuremath{\Gamma}} = 0.50 and Prandtl number Pr = 4.38}",
      journal = {Journal of Fluid Mechanics},
         year = 2011,
        month = jun,
       volume = {676},
        pages = {5-40},
          doi = {10.1017/S0022112010005963},
       adsurl = {https://ui.adsabs.harvard.edu/abs/2011JFM...676....5W}
}

@article{van_der_Poel_2011,
author={van der Poel, E. P.  and Stevens, R. J. and Lohse D.},
title={Connecting flow structures and heat flux in turbulent Rayleigh–Bénard convection},
journal={Physics Review E},
fjournal={Phys. Rev. E},
volume={84},
pages={045303},
year={2011},
}

@article{Kerr_periodicBC_1996,
author={Kerr, R.M.},
title={Rayleigh number scaling in numerical convection},
journal={Journal of Fluid Mechanics},
fjournal={J. Fluid Mech.},
volume={310},
pages={139-179},
year={1996},
}

@ARTICLE{Pandey_et_al_2021,
       author = {{Pandey}, Ambrish and {Schumacher}, J{\"o}rg and {Sreenivasan}, Katepalli R.},
        title = "{Non-Boussinesq Low-Prandtl-number Convection with a Temperature-dependent Thermal Diffusivity}",
      journal = {\apj},
         year = 2021,
        month = jan,
       volume = {907},
       number = {1},
          eid = {56},
        pages = {56},
          doi = {10.3847/1538-4357/abd1d8},
archivePrefix = {arXiv},
       eprint = {2010.11120},
 primaryClass = {physics.flu-dyn},
       adsurl = {https://ui.adsabs.harvard.edu/abs/2021ApJ...907...56P}
}

@ARTICLE{Ahlers_et_al_2009,
       author = {{Ahlers}, Guenter and {Grossmann}, Siegfried and {Lohse}, Detlef},
        title = "{Heat transfer and large scale dynamics in turbulent Rayleigh-B{\'e}nard convection}",
      journal = {Reviews of Modern Physics},
         year = 2009,
        month = apr,
       volume = {81},
       number = {2},
        pages = {503-537},
          doi = {10.1103/RevModPhys.81.503},
archivePrefix = {arXiv},
       eprint = {0811.0471},
 primaryClass = {physics.flu-dyn},
       adsurl = {https://ui.adsabs.harvard.edu/abs/2009RvMP...81..503A}
}

@article{Chilla_et_al_2012,
   author = {Chill\`{a}, F. and Schumacher, J.}, 
   title = {New perspectives in turbulent Rayleigh-B\'{e}nard convection},
   journal = {The European Physical Journal E},
   fjournal = {Eur. Phys. J. E},
   volume = {35},
   pages = {58},
   year={2012},
}

@ARTICLE{Benzi_et_al_1998,
       author = {{Benzi}, R. and {Toschi}, F. and {Tripiccione}, R.},
        title = "{On the Heat Transfer in Rayleigh B{\'e}nard Systems}",
      journal = {Journal of Statistical Physics},
         year = 1998,
        month = nov,
       volume = {93},
        pages = {901-918},
          doi = {10.1023/B:JOSS.0000033168.36971.59},
archivePrefix = {arXiv},
       eprint = {chao-dyn/9808020},
 primaryClass = {nlin.CD},
       adsurl = {https://ui.adsabs.harvard.edu/abs/1998JSP....93..901B}
}

@ARTICLE{Xi_et_al_2016,
       author = {{Xi}, Heng-Dong and {Zhang}, Yi-Bao and {Hao}, Jian-Tao and {Xia}, Ke-Qing},
        title = "{Higher-order flow modes in turbulent Rayleigh-B{\'e}nard convection}",
      journal = {Journal of Fluid Mechanics},
         year = 2016,
        month = oct,
       volume = {805},
        pages = {31-51},
          doi = {10.1017/jfm.2016.572},
       adsurl = {https://ui.adsabs.harvard.edu/abs/2016JFM...805...31X}
}

@ARTICLE{Pandey_Sreenivasan_EPL_2021,
       author = {{Pandey}, Ambrish and {Sreenivasan}, Katepalli R.},
        title = "{Convective heat transport in slender cells is close to that in wider cells at high Rayleigh and Prandtl numbers}",
      journal = {EPL (Europhysics Letters)},
         year = 2021,
        month = jul,
       volume = {135},
       number = {2},
          eid = {24001},
        pages = {24001},
          doi = {10.1209/0295-5075/ac1bc9},
archivePrefix = {arXiv},
       eprint = {2108.04593},
 primaryClass = {physics.flu-dyn},
       adsurl = {https://ui.adsabs.harvard.edu/abs/2021EL....13524001P}
}

@ARTICLE{Gallet_2012,
       author = {{Gallet}, B. and {Herault}, J. and {Laroche}, C. and {P{\'e}tr{\'e}lis}, F. and {Fauve}, S.},
        title = "{Reversals of a large-scale field generated over a turbulent background}",
      journal = {Geophysical and Astrophysical Fluid Dynamics},
         year = 2012,
        month = aug,
       volume = {106},
       number = {4-5},
        pages = {468-492},
          doi = {10.1080/03091929.2011.648629},
archivePrefix = {arXiv},
       eprint = {1102.0477},
 primaryClass = {physics.flu-dyn},
       adsurl = {https://ui.adsabs.harvard.edu/abs/2012GApFD.106..468G}
}

@ARTICLE{Christensen_Aubert_2006,
       author = {{Christensen}, U.~R. and {Aubert}, J.},
        title = "{Scaling properties of convection-driven dynamos in rotating spherical shells and application to planetary magnetic fields}",
      journal = {Geophysical Journal International},
         year = 2006,
        month = jul,
       volume = {166},
       number = {1},
        pages = {97-114},
          doi = {10.1111/j.1365-246X.2006.03009.x},
       adsurl = {https://ui.adsabs.harvard.edu/abs/2006GeoJI.166...97C}
}

@ARTICLE{Christensen_et_al_2009Natur,
       author = {{Christensen}, Ulrich R. and {Holzwarth}, Volkmar and {Reiners}, Ansgar},
        title = "{Energy flux determines magnetic field strength of planets and stars}",
      journal = {\nat},
         year = 2009,
        month = jan,
       volume = {457},
       number = {7226},
        pages = {167-169},
          doi = {10.1038/nature07626},
       adsurl = {https://ui.adsabs.harvard.edu/abs/2009Natur.457..167C}
}

@ARTICLE{Browning_2008ApJ,
       author = {{Browning}, Matthew K.},
        title = "{Simulations of Dynamo Action in Fully Convective Stars}",
      journal = {\apj},
         year = 2008,
        month = apr,
       volume = {676},
       number = {2},
        pages = {1262-1280},
          doi = {10.1086/527432},
archivePrefix = {arXiv},
       eprint = {0712.1603},
 primaryClass = {astro-ph},
       adsurl = {https://ui.adsabs.harvard.edu/abs/2008ApJ...676.1262B}
}

@ARTICLE{Kapyla_2021A&A,
       author = {{K{\"a}pyl{\"a}}, P.~J.},
        title = "{Star-in-a-box simulations of fully convective stars}",
      journal = {\aap},
         year = 2021,
        month = jul,
       volume = {651},
          eid = {A66},
        pages = {A66},
          doi = {10.1051/0004-6361/202040049},
archivePrefix = {arXiv},
       eprint = {2012.01259},
 primaryClass = {astro-ph.SR},
       adsurl = {https://ui.adsabs.harvard.edu/abs/2021A&A...651A..66K}
}

@ARTICLE{Bice_Toomre_2023ApJ,
       author = {{Bice}, Connor P. and {Toomre}, Juri},
        title = "{Nature of Intense Magnetism and Differential Rotation in Convective Dynamos of M-dwarf Stars with Tachoclines}",
      journal = {\apj},
         year = 2023,
        month = apr,
       volume = {947},
       number = {1},
          eid = {36},
        pages = {36},
          doi = {10.3847/1538-4357/acac78},
       adsurl = {https://ui.adsabs.harvard.edu/abs/2023ApJ...947...36B}
}

@ARTICLE{Chabrier_Kuker_2006,
       author = {{Chabrier}, G. and {K{\"u}ker}, M.},
        title = "{Large-scale {\ensuremath{\alpha}}\^2-dynamo in low-mass stars and brown dwarfs}",
      journal = {\aap},
         year = 2006,
        month = feb,
       volume = {446},
       number = {3},
        pages = {1027-1037},
          doi = {10.1051/0004-6361:20042475},
archivePrefix = {arXiv},
       eprint = {astro-ph/0510075},
 primaryClass = {astro-ph},
       adsurl = {https://ui.adsabs.harvard.edu/abs/2006A&A...446.1027C}
}

@ARTICLE{Shulyak_2017NatAs,
       author = {{Shulyak}, D. and {Reiners}, A. and {Engeln}, A. and {Malo}, L. and {Yadav}, R. and {Morin}, J. and {Kochukhov}, O.},
        title = "{Strong dipole magnetic fields in fast rotating fully convective stars}",
      journal = {Nature Astronomy},
         year = 2017,
        month = aug,
       volume = {1},
          eid = {0184},
        pages = {0184},
          doi = {10.1038/s41550-017-0184},
archivePrefix = {arXiv},
       eprint = {1801.08571},
 primaryClass = {astro-ph.SR},
       adsurl = {https://ui.adsabs.harvard.edu/abs/2017NatAs...1E.184S}
}

@ARTICLE{Gastine_Wicht_AA_2012,
       author = {{Gastine}, T. and {Duarte}, L. and {Wicht}, J.},
        title = "{Dipolar versus multipolar dynamos: the influence of the background density stratification}",
      journal = {\aap},
         year = 2012,
        month = oct,
       volume = {546},
          eid = {A19},
        pages = {A19},
          doi = {10.1051/0004-6361/201219799},
archivePrefix = {arXiv},
       eprint = {1208.6093},
 primaryClass = {astro-ph.SR},
       adsurl = {https://ui.adsabs.harvard.edu/abs/2012A&A...546A..19G}
}

@ARTICLE{Gastine_et_al_2013,
       author = {{Gastine}, T. and {Morin}, J. and {Duarte}, L. and {Reiners}, A. and {Christensen}, U.~R. and {Wicht}, J.},
        title = "{What controls the magnetic geometry of M dwarfs?}",
      journal = {\aap},
         year = 2013,
        month = jan,
       volume = {549},
          eid = {L5},
        pages = {L5},
          doi = {10.1051/0004-6361/201220317},
archivePrefix = {arXiv},
       eprint = {1212.0136},
 primaryClass = {astro-ph.SR},
       adsurl = {https://ui.adsabs.harvard.edu/abs/2013A&A...549L...5G}
}

@ARTICLE{Brun_2022ApJ,
       author = {{Brun}, Allan Sacha and {Strugarek}, Antoine and {Noraz}, Quentin and {Perri}, Barbara and {Varela}, Jacobo and {Augustson}, Kyle and {Charbonneau}, Paul and {Toomre}, Juri},
        title = "{Powering Stellar Magnetism: Energy Transfers in Cyclic Dynamos of Sun-like Stars}",
      journal = {\apj},
         year = 2022,
        month = feb,
       volume = {926},
       number = {1},
          eid = {21},
        pages = {21},
          doi = {10.3847/1538-4357/ac469b},
archivePrefix = {arXiv},
       eprint = {2201.13218},
 primaryClass = {astro-ph.SR},
       adsurl = {https://ui.adsabs.harvard.edu/abs/2022ApJ...926...21B}
}

@ARTICLE{Morin_etal_2010MNRAS,
       author = {{Morin}, J. and {Donati}, J. -F. and {Petit}, P. and {Delfosse}, X. and {Forveille}, T. and {Jardine}, M.~M.},
        title = "{Large-scale magnetic topologies of late M dwarfs*}",
      journal = {\mnras},
         year = 2010,
        month = oct,
       volume = {407},
       number = {4},
        pages = {2269-2286},
          doi = {10.1111/j.1365-2966.2010.17101.x},
archivePrefix = {arXiv},
       eprint = {1005.5552},
 primaryClass = {astro-ph.SR},
       adsurl = {https://ui.adsabs.harvard.edu/abs/2010MNRAS.407.2269M}
}

@ARTICLE{Kochukhov_2021AA,
       author = {{Kochukhov}, Oleg},
        title = "{Magnetic fields of M dwarfs}",
      journal = {\aapr},
         year = 2021,
        month = dec,
       volume = {29},
       number = {1},
          eid = {1},
        pages = {1},
          doi = {10.1007/s00159-020-00130-3},
archivePrefix = {arXiv},
       eprint = {2011.01781},
 primaryClass = {astro-ph.SR},
       adsurl = {https://ui.adsabs.harvard.edu/abs/2021A&ARv..29....1K}
}

@ARTICLE{Raedler_etal_1990AA,
       author = {{Raedler}, K. -H. and {Wiedemann}, E. and {Brandenburg}, A. and {Meinel}, R. and {Tuominen}, I.},
        title = "{Nonlinear mean-field dynamo models - Stability and evolution of three-dimensional magnetic field configurations.}",
      journal = {\aap},
         year = 1990,
        month = nov,
       volume = {239},
        pages = {413-423},
       adsurl = {https://ui.adsabs.harvard.edu/abs/1990A&A...239..413R}
}

@ARTICLE{Moss_Sokoloff_2009AA,
       author = {{Moss}, D. and {Sokoloff}, D.},
        title = "{Dynamo generated toroidal magnetic fields in rapidly rotating stars}",
      journal = {\aap},
         year = 2009,
        month = apr,
       volume = {497},
       number = {3},
        pages = {829-833},
          doi = {10.1051/0004-6361/200811426},
       adsurl = {https://ui.adsabs.harvard.edu/abs/2009A&A...497..829M}
}

@ARTICLE{Driscoll_Olson_2011,
       author = {{Driscoll}, P. and {Olson}, P.},
        title = "{Superchron cycles driven by variable core heat flow}",
      journal = {\grl},
         year = 2011,
        month = may,
       volume = {38},
       number = {9},
          eid = {L09304},
        pages = {L09304},
          doi = {10.1029/2011GL046808},
       adsurl = {https://ui.adsabs.harvard.edu/abs/2011GeoRL..38.9304D}
}

@ARTICLE{Glatzmaier_et_al_1999,
       author = {{Glatzmaier}, Gary A. and {Coe}, Robert S. and {Hongre}, Lionel and {Roberts}, Paul H.},
        title = "{The role of the Earth's mantle in controlling the frequency of geomagnetic reversals}",
      journal = {\nat},
         year = 1999,
        month = oct,
       volume = {401},
       number = {6756},
        pages = {885-890},
          doi = {10.1038/44776},
       adsurl = {https://ui.adsabs.harvard.edu/abs/1999Natur.401..885G}
}

@article{Olson_et_al_2010,
title = {Geodynamo reversal frequency and heterogeneous core–mantle boundary heat flow},
journal = {Physics of the Earth and Planetary Interiors},
volume = {180},
number = {1},
pages = {66-79},
year = {2010},
issn = {0031-9201},
doi = {https://doi.org/10.1016/j.pepi.2010.02.010},
url = {https://www.sciencedirect.com/science/article/pii/S0031920110000440},
author = {{Olson}, Peter L. and {Coe},Robert S. and {Driscoll}, Peter E. and {Glatzmaier}, Gary A. and {Roberts}, Paul H.}
}

@ARTICLE{Biggin_et_al_2012,
       author = {{Biggin}, A.~J. and {Steinberger}, B. and {Aubert}, J. and {Suttie}, N. and {Holme}, R. and {Torsvik}, T.~H. and {van der Meer}, D.~G. and {van Hinsbergen}, D.~J.~J.},
        title = "{Possible links between long-term geomagnetic variations and whole-mantle convection processes}",
      journal = {Nature Geoscience},
         year = 2012,
        month = aug,
       volume = {5},
       number = {8},
        pages = {526-533},
          doi = {10.1038/ngeo1521},
       adsurl = {https://ui.adsabs.harvard.edu/abs/2012NatGe...5..526B}
}

@ARTICLE{Valet_et_al_2005,
       author = {{Valet}, Jean-Pierre and {Meynadier}, Laure and {Guyodo}, Yohan},
        title = "{Geomagnetic dipole strength and reversal rate over the past two million years}",
      journal = {\nat},
         year = 2005,
        month = jun,
       volume = {435},
       number = {7043},
        pages = {802-805},
          doi = {10.1038/nature03674},
       adsurl = {https://ui.adsabs.harvard.edu/abs/2005Natur.435..802V}
}

@ARTICLE{Lay_et_al_2008,
       author = {{Lay}, Thorne and {Hernlund}, John and {Buffett}, Bruce A.},
        title = "{Core-mantle boundary heat flow}",
      journal = {Nature Geoscience},
         year = 2008,
        month = jan,
       volume = {1},
       number = {1},
        pages = {25-32},
          doi = {10.1038/ngeo.2007.44},
       adsurl = {https://ui.adsabs.harvard.edu/abs/2008NatGe...1...25L}
}

@ARTICLE{Wright_et_al_2011ApJ,
       author = {{Wright}, Nicholas J. and {Drake}, Jeremy J. and {Mamajek}, Eric E. and {Henry}, Gregory W.},
        title = "{The Stellar-activity-Rotation Relationship and the Evolution of Stellar Dynamos}",
      journal = {\apj},
         year = 2011,
        month = dec,
       volume = {743},
       number = {1},
          eid = {48},
        pages = {48},
          doi = {10.1088/0004-637X/743/1/48},
archivePrefix = {arXiv},
       eprint = {1109.4634},
 primaryClass = {astro-ph.SR},
       adsurl = {https://ui.adsabs.harvard.edu/abs/2011ApJ...743...48W}
}

@ARTICLE{Fan_Fang_2014ApJ,
       author = {{Fan}, Yuhong and {Fang}, Fang},
        title = "{A Simulation of Convective Dynamo in the Solar Convective Envelope: Maintenance of the Solar-like Differential Rotation and Emerging Flux}",
      journal = {\apj},
         year = 2014,
        month = jul,
       volume = {789},
       number = {1},
          eid = {35},
        pages = {35},
          doi = {10.1088/0004-637X/789/1/35},
archivePrefix = {arXiv},
       eprint = {1405.3926},
 primaryClass = {astro-ph.SR},
       adsurl = {https://ui.adsabs.harvard.edu/abs/2014ApJ...789...35F}
}

@ARTICLE{Karak_et_al_2015A&A,
       author = {{Karak}, B.~B. and {K{\"a}pyl{\"a}}, P.~J. and {K{\"a}pyl{\"a}}, M.~J. and {Brandenburg}, A. and {Olspert}, N. and {Pelt}, J.},
        title = "{Magnetically controlled stellar differential rotation near the transition from solar to anti-solar profiles}",
      journal = {\aap},
         year = 2015,
        month = apr,
       volume = {576},
          eid = {A26},
        pages = {A26},
          doi = {10.1051/0004-6361/201424521},
archivePrefix = {arXiv},
       eprint = {1407.0984},
 primaryClass = {astro-ph.SR},
       adsurl = {https://ui.adsabs.harvard.edu/abs/2015A&A...576A..26K}
}

@ARTICLE{Passos_et_al_2017A&A,
       author = {{Passos}, D. and {Miesch}, M. and {Guerrero}, G. and {Charbonneau}, P.},
        title = "{Meridional circulation dynamics in a cyclic convective dynamo}",
      journal = {\aap},
         year = 2017,
        month = nov,
       volume = {607},
          eid = {A120},
        pages = {A120},
          doi = {10.1051/0004-6361/201730568},
archivePrefix = {arXiv},
       eprint = {1702.02421},
 primaryClass = {astro-ph.SR},
       adsurl = {https://ui.adsabs.harvard.edu/abs/2017A&A...607A.120P}
}

@ARTICLE{Cossette_et_al_2017ApJ,
       author = {{Cossette}, Jean-Francois and {Charbonneau}, Paul and {Smolarkiewicz}, Piotr K. and {Rast}, Mark P.},
        title = "{Magnetically Modulated Heat Transport in a Global Simulation of Solar Magneto-convection}",
      journal = {\apj},
         year = 2017,
        month = may,
       volume = {841},
       number = {1},
          eid = {65},
        pages = {65},
          doi = {10.3847/1538-4357/aa6d60},
       adsurl = {https://ui.adsabs.harvard.edu/abs/2017ApJ...841...65C}
}

@ARTICLE{Kadanoff_et_al_1995PhFl,
       author = {{Kadanoff}, Leo and {Lohse}, Detlef and {Wang}, Jane and {Benzi}, Roberto},
        title = "{Scaling and dissipation in the GOY shell model}",
      journal = {Physics of Fluids},
         year = 1995,
        month = mar,
       volume = {7},
       number = {3},
        pages = {617-629},
          doi = {10.1063/1.868775},
archivePrefix = {arXiv},
       eprint = {chao-dyn/9409001},
 primaryClass = {nlin.CD},
       adsurl = {https://ui.adsabs.harvard.edu/abs/1995PhFl....7..617K}
}

@ARTICLE{Benzi_Pinton_2010PhRvL,
       author = {{Benzi}, Roberto and {Pinton}, Jean-Fran{\c{c}}ois},
        title = "{Magnetic Reversals in a Simple Model of Magnetohydrodynamics}",
      journal = {\prl},
         year = 2010,
        month = jul,
       volume = {105},
       number = {2},
          eid = {024501},
        pages = {024501},
          doi = {10.1103/PhysRevLett.105.024501},
archivePrefix = {arXiv},
       eprint = {0906.0427},
 primaryClass = {astro-ph.EP},
       adsurl = {https://ui.adsabs.harvard.edu/abs/2010PhRvL.105b4501B}
}

@ARTICLE{Benzi_Pinton_2011IJBC,
       author = {{Benzi}, Roberto and {Pinton}, Jean-Fran{\c{c}}ois},
        title = "{Stochastic Resonance in a Simple Model of Magnetic Reversals}",
      journal = {International Journal of Bifurcation and Chaos},
         year = 2011,
        month = jan,
       volume = {21},
       number = {12},
        pages = {3489},
          doi = {10.1142/S0218127411030684},
archivePrefix = {arXiv},
       eprint = {1104.4417},
 primaryClass = {nlin.CD},
       adsurl = {https://ui.adsabs.harvard.edu/abs/2011IJBC...21.3489B}
}

@ARTICLE{Kapyla_et_al_2009ApJ,
       author = {{K{\"a}pyl{\"a}}, Petri J. and {Korpi}, Maarit J. and {Brandenburg}, Axel},
        title = "{Large-scale Dynamos in Rigidly Rotating Turbulent Convection}",
      journal = {\apj},
         year = 2009,
        month = jun,
       volume = {697},
       number = {2},
        pages = {1153-1163},
          doi = {10.1088/0004-637X/697/2/1153},
archivePrefix = {arXiv},
       eprint = {0812.3958},
 primaryClass = {astro-ph},
       adsurl = {https://ui.adsabs.harvard.edu/abs/2009ApJ...697.1153K}
}

@ARTICLE{Kapyla_et_al_2010A&A,
       author = {{K{\"a}pyl{\"a}}, P.~J. and {Korpi}, M.~J. and {Brandenburg}, A.},
        title = "{Open and closed boundaries in large-scale convective dynamos}",
      journal = {\aap},
         year = 2010,
        month = jul,
       volume = {518},
          eid = {A22},
        pages = {A22},
          doi = {10.1051/0004-6361/200913722},
archivePrefix = {arXiv},
       eprint = {0911.4120},
 primaryClass = {astro-ph.SR},
       adsurl = {https://ui.adsabs.harvard.edu/abs/2010A&A...518A..22K}
}

@ARTICLE{Bushby_et_al_2018AA,
       author = {{Bushby}, P.~J. and {K{\"a}pyl{\"a}}, P.~J. and {Masada}, Y. and {Brandenburg}, A. and {Favier}, B. and {Guervilly}, C. and {K{\"a}pyl{\"a}}, M.~J.},
        title = "{Large-scale dynamos in rapidly rotating plane layer convection}",
      journal = {\aap},
         year = 2018,
        month = may,
       volume = {612},
          eid = {A97},
        pages = {A97},
          doi = {10.1051/0004-6361/201732066},
archivePrefix = {arXiv},
       eprint = {1710.03174},
 primaryClass = {astro-ph.SR},
       adsurl = {https://ui.adsabs.harvard.edu/abs/2018A&A...612A..97B}
}

@ARTICLE{Stauffer_2017AJ,
       author = {{Stauffer}, John and {Collier Cameron}, Andrew and {Jardine}, Moira and {David}, Trevor J. and {Rebull}, Luisa and {Cody}, Ann Marie and {Hillenbrand}, Lynne A. and {Barrado}, David and {Wolk}, Scott and {Davenport}, James and {Pinsonneault}, Marc},
        title = "{Orbiting Clouds of Material at the Keplerian Co-rotation Radius of Rapidly Rotating Low-mass WTTs in Upper Sco}",
      journal = {\aj},
         year = 2017,
        month = apr,
       volume = {153},
       number = {4},
          eid = {152},
        pages = {152},
          doi = {10.3847/1538-3881/aa5eb9},
archivePrefix = {arXiv},
       eprint = {1702.01797},
 primaryClass = {astro-ph.SR},
       adsurl = {https://ui.adsabs.harvard.edu/abs/2017AJ....153..152S}
}

@ARTICLE{Ryan_Sarson_2008EL,
       author = {{Ryan}, D.~A. and {Sarson}, G.~R.},
        title = "{The geodynamo as a low-dimensional deterministic system at the edge of chaos}",
      journal = {EPL (Europhysics Letters)},
         year = 2008,
        month = aug,
       volume = {83},
       number = {4},
        pages = {49001},
          doi = {10.1209/0295-5075/83/49001},
       adsurl = {https://ui.adsabs.harvard.edu/abs/2008EL.....8349001R}
}

@ARTICLE{Magaudda_et_al_2022,
       author = {{Magaudda}, E. and {Stelzer}, B. and {Raetz}, St.},
        title = "{First eROSITA-TESS results for M dwarfs: Mass dependence of the X-ray activity rotation relation and an assessment of sensitivity limits}",
      journal = {arXiv e-prints},
         year = 2022,
        month = jan,
          eid = {arXiv:2201.03897},
        pages = {arXiv:2201.03897},
archivePrefix = {arXiv},
       eprint = {2201.03897},
 primaryClass = {astro-ph.SR},
       adsurl = {https://ui.adsabs.harvard.edu/abs/2022arXiv220103897M}
}

@ARTICLE{Magaudda_et_al_2020A&A,
       author = {{Magaudda}, E. and {Stelzer}, B. and {Covey}, K.~R. and {Raetz}, St. and {Matt}, S.~P. and {Scholz}, A.},
        title = "{Relation of X-ray activity and rotation in M dwarfs and predicted time-evolution of the X-ray luminosity}",
      journal = {\aap},
         year = 2020,
        month = jun,
       volume = {638},
          eid = {A20},
        pages = {A20},
          doi = {10.1051/0004-6361/201937408},
archivePrefix = {arXiv},
       eprint = {2004.02904},
 primaryClass = {astro-ph.SR},
       adsurl = {https://ui.adsabs.harvard.edu/abs/2020A&A...638A..20M}
}

@ARTICLE{Pallavicini_et_al_1981ApJ,
       author = {{Pallavicini}, R. and {Golub}, L. and {Rosner}, R. and {Vaiana}, G.~S. and {Ayres}, T. and {Linsky}, J.~L.},
        title = "{Relations among stellar X-ray emission observed from Einstein, stellar rotation and bolometric luminosity.}",
      journal = {\apj},
         year = 1981,
        month = aug,
       volume = {248},
        pages = {279-290},
          doi = {10.1086/159152},
       adsurl = {https://ui.adsabs.harvard.edu/abs/1981ApJ...248..279P}
}

@ARTICLE{Pevtsov_et_al_2003ApJ,
       author = {{Pevtsov}, Alexei A. and {Fisher}, George H. and {Acton}, Loren W. and {Longcope}, Dana W. and {Johns-Krull}, Christopher M. and {Kankelborg}, Charles C. and {Metcalf}, Thomas R.},
        title = "{The Relationship Between X-Ray Radiance and Magnetic Flux}",
      journal = {\apj},
         year = 2003,
        month = dec,
       volume = {598},
       number = {2},
        pages = {1387-1391},
          doi = {10.1086/378944},
       adsurl = {https://ui.adsabs.harvard.edu/abs/2003ApJ...598.1387P}
}

@ARTICLE{Vilhu_Walter_1987ApJ,
       author = {{Vilhu}, Osmi and {Walter}, Frederick M.},
        title = "{Chromospheric-Coronal Activity at Saturated Levels}",
      journal = {\apj},
         year = 1987,
        month = oct,
       volume = {321},
        pages = {958},
          doi = {10.1086/165689},
       adsurl = {https://ui.adsabs.harvard.edu/abs/1987ApJ...321..958V}
}

@ARTICLE{Fleming_et_al_1995ApJ,
       author = {{Fleming}, Thomas A. and {Schmitt}, Juergen H.~M.~M. and {Giampapa}, Mark S.},
        title = "{Correlations of Coronal X-Ray Emission with Activity, Mass, and Age of the Nearby K and M Dwarfs}",
      journal = {\apj},
         year = 1995,
        month = sep,
       volume = {450},
        pages = {401},
          doi = {10.1086/176150},
       adsurl = {https://ui.adsabs.harvard.edu/abs/1995ApJ...450..401F}
}

@ARTICLE{Krishnamurthi_et_al_1998ApJ,
       author = {{Krishnamurthi}, Anita and {Terndrup}, D.~M. and {Pinsonneault}, M.~H. and {Sellgren}, K. and {Stauffer}, John R. and {Schild}, Rudolph and {Backman}, D.~E. and {Beisser}, K.~B. and {Dahari}, D.~B. and {Dasgupta}, A. and {Hagelgans}, J.~T. and {Seeds}, M.~A. and {Anand}, Rajan and {Laaksonen}, Bentley D. and {Marschall}, Laurence A. and {Ramseyer}, T.},
        title = "{New Rotation Periods in the Pleiades: Interpreting Activity Indicators}",
      journal = {\apj},
         year = 1998,
        month = jan,
       volume = {493},
       number = {2},
        pages = {914-925},
          doi = {10.1086/305173},
archivePrefix = {arXiv},
       eprint = {astro-ph/9711284},
 primaryClass = {astro-ph},
       adsurl = {https://ui.adsabs.harvard.edu/abs/1998ApJ...493..914K}
}

@ARTICLE{Charbonneau_2020LRSP,
       author = {{Charbonneau}, Paul},
        title = "{Dynamo models of the solar cycle}",
      journal = {Living Reviews in Solar Physics},
         year = 2020,
        month = dec,
       volume = {17},
       number = {1},
          eid = {4},
        pages = {4},
          doi = {10.1007/s41116-020-00025-6},
       adsurl = {https://ui.adsabs.harvard.edu/abs/2020LRSP...17....4C}
}

@ARTICLE{Nigro_2019NCimC,
       author = {{Nigro}, G.},
        title = "{Attracted by the fascinating magnetism of the Sun}",
      journal = {Nuovo Cimento C Geophysics Space Physics C},
         year = 2019,
        month = jan,
       volume = {42},
       number = {1},
          eid = {2},
        pages = {2},
          doi = {10.1393/ncc/i2019-19002-5},
       adsurl = {https://ui.adsabs.harvard.edu/abs/2019NCimC..42....2N}
}

@ARTICLE{Charbonneau_MacGregor_1996ApJ,
       author = {{Charbonneau}, P. and {MacGregor}, K.~B.},
        title = "{On the Generation of Equipartition-Strength Magnetic Fields by Turbulent Hydromagnetic Dynamos}",
      journal = {\apjl},
         year = 1996,
        month = dec,
       volume = {473},
        pages = {L59},
          doi = {10.1086/310387},
       adsurl = {https://ui.adsabs.harvard.edu/abs/1996ApJ...473L..59C}
}

@ARTICLE{Archontis_et_al_2007AA,
       author = {{Archontis}, V. and {Dorch}, S.~B.~F. and {Nordlund}, {\r{A}}.},
        title = "{Nonlinear MHD dynamo operating at equipartition}",
      journal = {\aap},
         year = 2007,
        month = sep,
       volume = {472},
       number = {3},
        pages = {715-726},
          doi = {10.1051/0004-6361:20065087},
       adsurl = {https://ui.adsabs.harvard.edu/abs/2007A&A...472..715A}
}

@ARTICLE{Tobias_2021JFM,
       author = {{Tobias}, S.~M.},
        title = "{The turbulent dynamo}",
      journal = {Journal of Fluid Mechanics},
         year = 2021,
        month = apr,
       volume = {912},
          eid = {P1},
        pages = {P1},
          doi = {10.1017/jfm.2020.1055},
archivePrefix = {arXiv},
       eprint = {1907.03685},
 primaryClass = {physics.flu-dyn},
       adsurl = {https://ui.adsabs.harvard.edu/abs/2021JFM...912P...1T}
}

@ARTICLE{Biermann_et_al_1951PhRv,
       author = {{Biermann}, Ludwig and {Schl{\"u}ter}, Arnulf},
        title = "{Cosmic Radiation and Cosmic Magnetic Fields. II. Origin of Cosmic Magnetic Fields}",
      journal = {Physical Review},
         year = 1951,
        month = jun,
       volume = {82},
       number = {6},
        pages = {863-868},
          doi = {10.1103/PhysRev.82.863},
       adsurl = {https://ui.adsabs.harvard.edu/abs/1951PhRv...82..863B}
}

@ARTICLE{Kulsrud_et_al_1997ApJ,
       author = {{Kulsrud}, Russell M. and {Cen}, Renyue and {Ostriker}, Jeremiah P. and {Ryu}, Dongsu},
        title = "{The Protogalactic Origin for Cosmic Magnetic Fields}",
      journal = {\apj},
         year = 1997,
        month = may,
       volume = {480},
       number = {2},
        pages = {481-491},
          doi = {10.1086/303987},
archivePrefix = {arXiv},
       eprint = {astro-ph/9607141},
 primaryClass = {astro-ph},
       adsurl = {https://ui.adsabs.harvard.edu/abs/1997ApJ...480..481K}
}

@ARTICLE{Brandenburg_Subramanian_2000A&A,
       author = {{Brandenburg}, A. and {Subramanian}, K.},
        title = "{Large scale dynamos with ambipolar diffusion nonlinearity}",
      journal = {\aap},
         year = 2000,
        month = sep,
       volume = {361},
        pages = {L33-L36},
archivePrefix = {arXiv},
       eprint = {astro-ph/0007450},
 primaryClass = {astro-ph},
       adsurl = {https://ui.adsabs.harvard.edu/abs/2000A&A...361L..33B}
}

@ARTICLE{Cho_et_al_2009ApJ,
       author = {{Cho}, Jungyeon and {Vishniac}, Ethan T. and {Beresnyak}, Andrey and {Lazarian}, A. and {Ryu}, Dongsu},
        title = "{Growth of Magnetic Fields Induced by Turbulent Motions}",
      journal = {\apj},
         year = 2009,
        month = mar,
       volume = {693},
       number = {2},
        pages = {1449-1461},
          doi = {10.1088/0004-637X/693/2/1449},
archivePrefix = {arXiv},
       eprint = {0812.0817},
 primaryClass = {astro-ph},
       adsurl = {https://ui.adsabs.harvard.edu/abs/2009ApJ...693.1449C}
}

@ARTICLE{Beresnyak_2012PhRvL,
       author = {{Beresnyak}, A.},
        title = "{Universal Nonlinear Small-Scale Dynamo}",
      journal = {\prl},
         year = 2012,
        month = jan,
       volume = {108},
       number = {3},
          eid = {035002},
        pages = {035002},
          doi = {10.1103/PhysRevLett.108.035002},
archivePrefix = {arXiv},
       eprint = {1109.4644},
 primaryClass = {astro-ph.GA},
       adsurl = {https://ui.adsabs.harvard.edu/abs/2012PhRvL.108c5002B}
}

@ARTICLE{Tzeferacos_2018NatCo,
       author = {{Tzeferacos}, P. and {Rigby}, A. and {Bott}, A.~F.~A. and {Bell}, A.~R. and {Bingham}, R. and {Casner}, A. and {Cattaneo}, F. and {Churazov}, E.~M. and {Emig}, J. and {Fiuza}, F. and {Forest}, C.~B. and {Foster}, J. and {Graziani}, C. and {Katz}, J. and {Koenig}, M. and {Li}, C. -K. and {Meinecke}, J. and {Petrasso}, R. and {Park}, H. -S. and {Remington}, B.~A. and {Ross}, J.~S. and {Ryu}, D. and {Ryutov}, D. and {White}, T.~G. and {Reville}, B. and {Miniati}, F. and {Schekochihin}, A.~A. and {Lamb}, D.~Q. and {Froula}, D.~H. and {Gregori}, G.},
        title = "{Laboratory evidence of dynamo amplification of magnetic fields in a turbulent plasma}",
      journal = {Nature Communications},
         year = 2018,
        month = feb,
       volume = {9},
          eid = {591},
        pages = {591},
          doi = {10.1038/s41467-018-02953-2},
archivePrefix = {arXiv},
       eprint = {1702.03016},
 primaryClass = {physics.plasm-ph},
       adsurl = {https://ui.adsabs.harvard.edu/abs/2018NatCo...9..591T}
}

@ARTICLE{Sorriso_et_al_2019FrP,
       author = {{Sorriso-Valvo}, Luca and {De Vita}, Gaetano and {Fraternale}, Federico and {Gurchumelia}, Alexandre and {Perri}, Silvia and {Nigro}, Giuseppina and {Catapano}, Filomena and {Retin{\`o}}, Alessandro and {Chen}, Christopher H.~K. and {Yordanova}, Emiliya and {Pezzi}, Oreste and {Chargazia}, Khatuna and {Kharshiladze}, Oleg and {Kvaratskhelia}, Diana and {V{\'a}sconez}, Christian L. and {Marino}, Raffaele and {Le Contel}, Olivier and {Giles}, Barbara and {Moore}, Thomas E. and {Torbert}, Roy B. and {Burch}, James L.},
        title = "{Sign singularity of the local energy transfer in space plasma turbulence}",
      journal = {Frontiers in Physics},
         year = 2019,
        month = aug,
       volume = {7},
          eid = {108},
        pages = {108},
          doi = {10.3389/fphy.2019.00108},
archivePrefix = {arXiv},
       eprint = {1907.11108},
 primaryClass = {physics.space-ph},
       adsurl = {https://ui.adsabs.harvard.edu/abs/2019FrP.....7..108S}
}

@ARTICLE{Cadavid_et_al_1994ApJ,
       author = {{Cadavid}, A.~C. and {Lawrence}, J.~K. and {Ruzmaikin}, A.~A. and {Kayleng-Knight}, A.},
        title = "{Multifractal Models of Small-Scale Solar Magnetic Fields}",
      journal = {\apj},
         year = 1994,
        month = jul,
       volume = {429},
        pages = {391},
          doi = {10.1086/174329},
       adsurl = {https://ui.adsabs.harvard.edu/abs/1994ApJ...429..391C}
}

@ARTICLE{Cattaneo_Tobia_2005PhFl,
       author = {{Cattaneo}, Fausto and {Tobias}, Steven M.},
        title = "{Interaction between dynamos at different scales}",
      journal = {Physics of Fluids},
         year = 2005,
        month = dec,
       volume = {17},
       number = {12},
          eid = {127105-127105-6},
        pages = {127105-127105-6},
          doi = {10.1063/1.2145755},
       adsurl = {https://ui.adsabs.harvard.edu/abs/2005PhFl...17l7105C}
}

@ARTICLE{Martin_et_al_2013PhRvE,
       author = {{Martin}, L.~N. and {De Vita}, G. and {Sorriso-Valvo}, L. and {Dmitruk}, P. and {Nigro}, G. and {Primavera}, L. and {Carbone}, V.},
        title = "{Cancellation properties in Hall magnetohydrodynamics with a strong guide magnetic field}",
      journal = {\pre},
         year = 2013,
        month = dec,
       volume = {88},
       number = {6},
          eid = {063107},
        pages = {063107},
          doi = {10.1103/PhysRevE.88.063107},
       adsurl = {https://ui.adsabs.harvard.edu/abs/2013PhRvE..88f3107M}
}

@ARTICLE{Munoz_et_al_2021,
       author = {{Mu{\~n}oz}, V{\'\i}ctor and {Dom{\'\i}nguez}, Macarena and {Nigro}, Giuseppina and {Riquelme}, Mario and {Carbone}, Vincenzo},
        title = "{Fractality of an MHD shell model for turbulent plasma driven by solar wind data: A review}",
      journal = {Journal of Atmospheric and Solar-Terrestrial Physics},
         year = 2021,
        month = mar,
       volume = {214},
          eid = {105524},
        pages = {105524},
          doi = {10.1016/j.jastp.2020.105524},
       adsurl = {https://ui.adsabs.harvard.edu/abs/2021JASTP.21405524M}
}

@ARTICLE{Yamada_Ohkitani_1988PRL,
       author = {{Yamada}, Michio and {Ohkitani}, Koji},
        title = "{Lyapunov spectrum of a model of two-dimensional turbulence}",
      journal = {\prl},
         year = 1988,
        month = mar,
       volume = {60},
       number = {11},
        pages = {983-986},
          doi = {10.1103/PhysRevLett.60.983},
       adsurl = {https://ui.adsabs.harvard.edu/abs/1988PhRvL..60..983Y}
}

@ARTICLE{Mascareno_et_al_2016A&A,
       author = {{Su{\'a}rez Mascare{\~n}o}, A. and {Rebolo}, R. and {Gonz{\'a}lez Hern{\'a}ndez}, J.~I.},
        title = "{Magnetic cycles and rotation periods of late-type stars from photometric time series}",
      journal = {\aap},
         year = 2016,
        month = oct,
       volume = {595},
          eid = {A12},
        pages = {A12},
          doi = {10.1051/0004-6361/201628586},
archivePrefix = {arXiv},
       eprint = {1607.03049},
 primaryClass = {astro-ph.SR},
       adsurl = {https://ui.adsabs.harvard.edu/abs/2016A&A...595A..12S}
}

@ARTICLE{Wargelin_et_al_2017MNRAS,
       author = {{Wargelin}, B.~J. and {Saar}, S.~H. and {Pojma{\'n}ski}, G. and {Drake}, J.~J. and {Kashyap}, V.~L.},
        title = "{Optical, UV, and X-ray evidence for a 7-yr stellar cycle in Proxima Centauri}",
      journal = {\mnras},
         year = 2017,
        month = jan,
       volume = {464},
       number = {3},
        pages = {3281-3296},
          doi = {10.1093/mnras/stw2570},
archivePrefix = {arXiv},
       eprint = {1610.03447},
 primaryClass = {astro-ph.SR},
       adsurl = {https://ui.adsabs.harvard.edu/abs/2017MNRAS.464.3281W}
}

@ARTICLE{Jeffers_et_al_2022A&A,
       author = {{Jeffers}, S.~V. and {Cameron}, R.~H. and {Marsden}, S.~C. and {Boro Saikia}, S. and {Folsom}, C.~P. and {Jardine}, M.~M. and {Morin}, J. and {Petit}, P. and {See}, V. and {Vidotto}, A.~A. and {Wolter}, U. and {Mittag}, M.},
        title = "{The crucial role of surface magnetic fields for stellar dynamos: ϵ Eridani, 61 Cygni A, and the Sun}",
      journal = {\aap},
         year = 2022,
        month = may,
       volume = {661},
          eid = {A152},
        pages = {A152},
          doi = {10.1051/0004-6361/202142202},
archivePrefix = {arXiv},
       eprint = {2201.07530},
 primaryClass = {astro-ph.SR},
       adsurl = {https://ui.adsabs.harvard.edu/abs/2022A&A...661A.152J}
}

@ARTICLE{Metcalfe_et_al_2013ApJ,
       author = {{Metcalfe}, T.~S. and {Buccino}, A.~P. and {Brown}, B.~P. and {Mathur}, S. and {Soderblom}, D.~R. and {Henry}, T.~J. and {Mauas}, P.~J.~D. and {Petrucci}, R. and {Hall}, J.~C. and {Basu}, S.},
        title = "{Magnetic Activity Cycles in the Exoplanet Host Star epsilon Eridani}",
      journal = {\apjl},
         year = 2013,
        month = feb,
       volume = {763},
       number = {2},
          eid = {L26},
        pages = {L26},
          doi = {10.1088/2041-8205/763/2/L26},
archivePrefix = {arXiv},
       eprint = {1212.4425},
 primaryClass = {astro-ph.SR},
       adsurl = {https://ui.adsabs.harvard.edu/abs/2013ApJ...763L..26M}
}

@ARTICLE{Ruiz_Nelson_1981PhRvA,
       author = {{Ruiz}, R. and {Nelson}, D.~R.},
        title = "{Turbulence in binary fluid mixtures}",
      journal = {\pra},
         year = 1981,
        month = jun,
       volume = {23},
       number = {6},
        pages = {3224-3246},
          doi = {10.1103/PhysRevA.23.3224},
       adsurl = {https://ui.adsabs.harvard.edu/abs/1981PhRvA..23.3224R}
}

@ARTICLE{Yamada_Ohkitani_1988PThPh,
       author = {{Yamada}, M. and {Ohkitani}, K.},
        title = "{The Inertial Subrange and Non-Positive Lyapunov Exponents in Fully-Developed Turbulence}",
      journal = {Progress of Theoretical Physics},
         year = 1988,
        month = jun,
       volume = {79},
       number = {6},
        pages = {1265-1268},
          doi = {10.1143/PTP.79.1265},
       adsurl = {https://ui.adsabs.harvard.edu/abs/1988PThPh..79.1265Y}
}

@ARTICLE{Verma_Kumar_2016JTurb,
       author = {{Verma}, Mahendra K. and {Kumar}, Rohit},
        title = "{Dynamos at extreme magnetic Prandtl numbers: insights from shell models}",
      journal = {Journal of Turbulence},
         year = 2016,
        month = dec,
       volume = {17},
       number = {12},
        pages = {1112-1141},
          doi = {10.1080/14685248.2016.1228937},
archivePrefix = {arXiv},
       eprint = {1510.00122},
 primaryClass = {physics.flu-dyn},
       adsurl = {https://ui.adsabs.harvard.edu/abs/2016JTurb..17.1112V}
}

@ARTICLE{Nigro_2022,
       author = {{Nigro}, Giuseppina},
        title = "{An Argument in Favor of Magnetic Polarity Reversals Due to Heat Flux Variations in Fully Convective Stars and Planets}",
      journal = {\apj},
         year = 2022,
        month = oct,
       volume = {938},
       number = {1},
          eid = {22},
        pages = {22},
          doi = {10.3847/1538-4357/ac8d57},
       adsurl = {https://ui.adsabs.harvard.edu/abs/2022ApJ...938...22N}
}

@ARTICLE{Sorriso_et_al_2007PEPI,
       author = {{Sorriso-Valvo}, Luca and {Stefani}, Frank and {Carbone}, Vincenzo and {Nigro}, Giuseppina and {Lepreti}, Fabio and {Vecchio}, Antonio and {Veltri}, Pierluigi},
        title = "{A statistical analysis of polarity reversals of the geomagnetic field}",
      journal = {Physics of the Earth and Planetary Interiors},
         year = 2007,
        month = oct,
       volume = {164},
       number = {3-4},
        pages = {197-207},
          doi = {10.1016/j.pepi.2007.07.001},
       adsurl = {https://ui.adsabs.harvard.edu/abs/2007PEPI..164..197S}
}

@ARTICLE{Jonkers_2003PEPI,
       author = {{Jonkers}, A.~R.~T.},
        title = "{Long-range dependence in the Cenozoic reversal record}",
      journal = {Physics of the Earth and Planetary Interiors},
         year = 2003,
        month = mar,
       volume = {135},
       number = {4},
        pages = {253-266},
          doi = {10.1016/S0031-9201(03)00036-0},
       adsurl = {https://ui.adsabs.harvard.edu/abs/2003PEPI..135..253J}
}

@ARTICLE{Olson_et_al_2014E&PSL,
       author = {{Olson}, Peter and {Hinnov}, Linda A. and {Driscoll}, Peter E.},
        title = "{Nonrandom geomagnetic reversal times and geodynamo evolution}",
      journal = {Earth and Planetary Science Letters},
         year = 2014,
        month = feb,
       volume = {388},
        pages = {9-17},
          doi = {10.1016/j.epsl.2013.11.038},
       adsurl = {https://ui.adsabs.harvard.edu/abs/2014E&PSL.388....9O}
}

@ARTICLE{Ryan_Sarson_2011PEPI,
       author = {{Ryan}, David A. and {Sarson}, Graeme R.},
        title = "{A coupled low order dynamo/turbulent shell model for geomagnetic field variations and reversals}",
      journal = {Physics of the Earth and Planetary Interiors},
         year = 2011,
        month = oct,
       volume = {188},
       number = {3},
        pages = {214-234},
          doi = {10.1016/j.pepi.2011.09.003},
       adsurl = {https://ui.adsabs.harvard.edu/abs/2011PEPI..188..214R}
}

@ARTICLE{McFadden_Merrill_1984JGR,
       author = {{McFadden}, P.~L. and {Merrill}, R.~T.},
        title = "{Lower mantle convection and geomagnetism}",
      journal = {\jgr},
         year = 1984,
        month = may,
       volume = {89},
       number = {B5},
        pages = {3354-3362},
          doi = {10.1029/JB089iB05p03354},
       adsurl = {https://ui.adsabs.harvard.edu/abs/1984JGR....89.3354M}
}

@ARTICLE{Morin_et_al_2011MNRAS,
       author = {{Morin}, J. and {Dormy}, E. and {Schrinner}, M. and {Donati}, J. -F.},
        title = "{Weak- and strong-field dynamos: from the Earth to the stars}",
      journal = {\mnras},
         year = 2011,
        month = nov,
       volume = {418},
       number = {1},
        pages = {L133-L137},
          doi = {10.1111/j.1745-3933.2011.01159.x},
archivePrefix = {arXiv},
       eprint = {1106.4263},
 primaryClass = {astro-ph.SR},
       adsurl = {https://ui.adsabs.harvard.edu/abs/2011MNRAS.418L.133M}
}

@ARTICLE{Kitchatinov_et_al_2014MNRAS,
       author = {{Kitchatinov}, L.~L. and {Moss}, D. and {Sokoloff}, D.},
        title = "{Magnetic fields in fully convective M-dwarfs: oscillatory dynamos versus bistability.}",
      journal = {\mnras},
         year = 2014,
        month = jul,
       volume = {442},
        pages = {L1-L4},
          doi = {10.1093/mnrasl/slu041},
archivePrefix = {arXiv},
       eprint = {1401.1764},
 primaryClass = {astro-ph.SR},
       adsurl = {https://ui.adsabs.harvard.edu/abs/2014MNRAS.442L...1K}
}

@ARTICLE{Wright_Drake_2016Natur,
       author = {{Wright}, Nicholas J. and {Drake}, Jeremy J.},
        title = "{Solar-type dynamo behaviour in fully convective stars without a tachocline}",
      journal = {\nat},
         year = 2016,
        month = jul,
       volume = {535},
       number = {7613},
        pages = {526-528},
          doi = {10.1038/nature18638},
archivePrefix = {arXiv},
       eprint = {1607.07870},
 primaryClass = {astro-ph.SR},
       adsurl = {https://ui.adsabs.harvard.edu/abs/2016Natur.535..526W}
}

@article{LEO,
   author = {{Primavera}, Leonardo and {Florio}, Emilia}, 
   title = {Parallel Algorithms for Multifractal Analysis of River Networks},
   journal = {Lecture Notes in Computer Science, including subseries Lecture Notes in Artificial Intelligence and
Lecture Notes in Bioinformatics},
   fjournal = {},
   volume = {11973},
   pages = {307–317},
   year={2020},
}

@ARTICLE{Yadav_et_al_2015ApJ,
       author = {{Yadav}, Rakesh K. and {Christensen}, Ulrich R. and {Morin}, Julien and {Gastine}, Thomas and {Reiners}, Ansgar and {Poppenhaeger}, Katja and {Wolk}, Scott J.},
        title = "{Explaining the Coexistence of Large-scale and Small-scale Magnetic Fields in Fully Convective Stars}",
      journal = {\apjl},
         year = 2015,
        month = nov,
       volume = {813},
       number = {2},
          eid = {L31},
        pages = {L31},
          doi = {10.1088/2041-8205/813/2/L31},
archivePrefix = {arXiv},
       eprint = {1510.05541},
 primaryClass = {astro-ph.SR},
       adsurl = {https://ui.adsabs.harvard.edu/abs/2015ApJ...813L..31Y}
}

@ARTICLE{Irving_Saar_et_al_2023ApJ,
       author = {{Irving}, Zackery A. and {Saar}, Steven H. and {Wargelin}, Bradford J. and {do Nascimento}, Jos{\'e}-Dias},
        title = "{Stellar Cycles in Fully Convective Stars and a New Interpretation of Dynamo Evolution}",
      journal = {\apj},
         year = 2023,
        month = jun,
       volume = {949},
       number = {2},
          eid = {51},
        pages = {51},
          doi = {10.3847/1538-4357/acc468},
archivePrefix = {arXiv},
       eprint = {2303.08519},
 primaryClass = {astro-ph.SR},
       adsurl = {https://ui.adsabs.harvard.edu/abs/2023ApJ...949...51I}
}

@ARTICLE{Route_2016ApJ,
       author = {{Route}, Matthew},
        title = "{The Discovery of Solar-like Activity Cycles Beyond the End of the Main Sequence?}",
      journal = {\apjl},
         year = 2016,
        month = oct,
       volume = {830},
       number = {2},
          eid = {L27},
        pages = {L27},
          doi = {10.3847/2041-8205/830/2/L27},
archivePrefix = {arXiv},
       eprint = {1609.07761},
 primaryClass = {astro-ph.SR},
       adsurl = {https://ui.adsabs.harvard.edu/abs/2016ApJ...830L..27R}
}

@ARTICLE{Hubbard_Brandenburg_2010GApFD,
       author = {{Hubbard}, Alexander and {Brandenburg}, Axel},
        title = "{Magnetic helicity fluxes in an {\ensuremath{\alpha}}2 dynamo embedded in a halo}",
      journal = {Geophysical and Astrophysical Fluid Dynamics},
         year = 2010,
        month = oct,
       volume = {104},
       number = {5},
        pages = {577-590},
          doi = {10.1080/03091929.2010.506438},
archivePrefix = {arXiv},
       eprint = {1004.4591},
 primaryClass = {astro-ph.SR},
       adsurl = {https://ui.adsabs.harvard.edu/abs/2010GApFD.104..577H}
}

@ARTICLE{Del_Sordo_et_al_2013MNRAS,
       author = {{Del Sordo}, F. and {Guerrero}, G. and {Brandenburg}, A.},
        title = "{Turbulent dynamos with advective magnetic helicity flux}",
      journal = {\mnras},
         year = 2013,
        month = feb,
       volume = {429},
       number = {2},
        pages = {1686-1694},
          doi = {10.1093/mnras/sts398},
archivePrefix = {arXiv},
       eprint = {1205.3502},
 primaryClass = {astro-ph.GA},
       adsurl = {https://ui.adsabs.harvard.edu/abs/2013MNRAS.429.1686D}
}

@ARTICLE{Brandenburg_2018JPlPh,
       author = {{Brandenburg}, Axel},
        title = "{Advances in mean-field dynamo theory and applications to astrophysical turbulence}",
      journal = {Journal of Plasma Physics},
         year = 2018,
        month = aug,
       volume = {84},
       number = {4},
          eid = {735840404},
        pages = {735840404},
          doi = {10.1017/S0022377818000806},
archivePrefix = {arXiv},
       eprint = {1801.05384},
 primaryClass = {physics.flu-dyn},
       adsurl = {https://ui.adsabs.harvard.edu/abs/2018JPlPh..84d7304B}
}

@ARTICLE{Cattaneo_Bodo_Tobias_2020_2020JPlPh,
       author = {{Cattaneo}, F. and {Bodo}, G. and {Tobias}, S.M},
        title = "{On magnetic helicity generation and transport in a nonlinear dynamo driven by a helical flow}",
      journal = {Journal of Plasma Physics},
         year = 2020,
        month = aug,
       volume = {86},
       number = {4},
          eid = {905860408},
        pages = {905860408},
          doi = {10.1017/S0022377820000690},
archivePrefix = {arXiv},
       eprint = {1801.05384},
 primaryClass = {physics.flu-dyn},
       adsurl =  {https:https://ui.adsabs.harvard.edu/abs/10.1017/S0022377820000690}
}

@ARTICLE{Dubrulle_et_al_2007NJPh,
       author = {{Dubrulle}, B. and {Blaineau}, P. and {Mafra Lopes}, O. and {Daviaud}, F. and {Laval}, J. -P. and {Dolganov}, R.},
        title = "{Bifurcations and dynamo action in a Taylor Green flow}",
      journal = {New Journal of Physics},
         year = 2007,
        month = aug,
       volume = {9},
       number = {8},
        pages = {308},
          doi = {10.1088/1367-2630/9/8/308},
       adsurl = {https://ui.adsabs.harvard.edu/abs/2007NJPh....9..308D}
}

@ARTICLE{Verma_et_al_2008PhRvE,
       author = {{Verma}, Mahendra K. and {Lessinnes}, Thomas and {Carati}, Daniele and {Sarris}, Ioannis and {Kumar}, Krishna and {Singh}, Meenakshi},
        title = "{Dynamo transition in low-dimensional models}",
      journal = {\pre},
         year = 2008,
        month = sep,
       volume = {78},
       number = {3},
          eid = {036409},
        pages = {036409},
          doi = {10.1103/PhysRevE.78.036409},
archivePrefix = {arXiv},
       eprint = {0801.2656},
 primaryClass = {nlin.CD},
       adsurl = {https://ui.adsabs.harvard.edu/abs/2008PhRvE..78c6409V}
}

@ARTICLE{Perrone_et_al_2011ApJ,
       author = {{Perrone}, D. and {Nigro}, G. and {Veltri}, P.},
        title = "{A Shell Model Turbulent Dynamo}",
      journal = {\apj},
         year = 2011,
        month = jul,
       volume = {735},
       number = {2},
          eid = {73},
        pages = {73},
          doi = {10.1088/0004-637X/735/2/73},
       adsurl = {https://ui.adsabs.harvard.edu/abs/2011ApJ...735...73P}
}

@ARTICLE{Field_Blackman_2002ApJ,
       author = {{Field}, George B. and {Blackman}, Eric G.},
        title = "{Dynamical Quenching of the {\ensuremath{\alpha}}$^{2}$ Dynamo}",
      journal = {\apj},
         year = 2002,
        month = jun,
       volume = {572},
       number = {1},
        pages = {685-692},
          doi = {10.1086/340233},
archivePrefix = {arXiv},
       eprint = {astro-ph/0111470},
 primaryClass = {astro-ph},
       adsurl = {https://ui.adsabs.harvard.edu/abs/2002ApJ...572..685F}
}

@ARTICLE{Gubbins_2008Natur,
       author = {{Gubbins}, David},
        title = "{Earth science: Geomagnetic reversals}",
      journal = {\nat},
         year = 2008,
        month = mar,
       volume = {452},
       number = {7184},
        pages = {165-167},
          doi = {10.1038/452165a},
       adsurl = {https://ui.adsabs.harvard.edu/abs/2008Natur.452..165G}
}

@ARTICLE{Shcherbakov_Fabian_2012JGRB,
       author = {{Shcherbakov}, Valeriy and {Fabian}, Karl},
        title = "{The geodynamo as a random walker: A view on reversal statistics}",
      journal = {Journal of Geophysical Research (Solid Earth)},
         year = 2012,
        month = mar,
       volume = {117},
       number = {B3},
          eid = {B03101},
        pages = {B03101},
          doi = {10.1029/2011JB008931},
       adsurl = {https://ui.adsabs.harvard.edu/abs/2012JGRB..117.3101S}
}

@ARTICLE{Constable_2000PEPI,
       author = {{Constable}, Catherine},
        title = "{On rates of occurrence of geomagnetic reversals}",
      journal = {Physics of the Earth and Planetary Interiors},
         year = 2000,
        month = mar,
       volume = {118},
       number = {3-4},
        pages = {181-193},
          doi = {10.1016/S0031-9201(99)00139-9},
       adsurl = {https://ui.adsabs.harvard.edu/abs/2000PEPI..118..181C}
}

@ARTICLE{Davenport_et_al_AJ_2020,
       author = {{Davenport}, James. R.~A. and {Mendoza}, Guadalupe Tovar and {Hawley}, Suzanne L.},
        title = "{10 Years of Stellar Activity for GJ 1243}",
      journal = {\aj},
         year = 2020,
        month = jul,
       volume = {160},
       number = {1},
          eid = {36},
        pages = {36},
          doi = {10.3847/1538-3881/ab9536},
archivePrefix = {arXiv},
       eprint = {2005.10281},
 primaryClass = {astro-ph.SR},
       adsurl = {https://ui.adsabs.harvard.edu/abs/2020AJ....160...36D}
}

@ARTICLE{Donati_et_al_Sci_2006,
       author = {{Donati}, Jean-Fran{\c{c}}ois and {Forveille}, Thierry and {Collier Cameron}, Andrew and {Barnes}, John R. and {Delfosse}, Xavier and {Jardine}, Moira M. and {Valenti}, Jeff A.},
        title = "{The Large-Scale Axisymmetric Magnetic Topology of a Very-Low-Mass Fully Convective Star}",
      journal = {Science},
         year = 2006,
        month = feb,
       volume = {311},
       number = {5761},
        pages = {633-635},
          doi = {10.1126/science.1121102},
archivePrefix = {arXiv},
       eprint = {astro-ph/0602069},
 primaryClass = {astro-ph},
       adsurl = {https://ui.adsabs.harvard.edu/abs/2006Sci...311..633D}
}

@ARTICLE{Morin_et_al_2008MNRAS_a,
       author = {{Morin}, J. and {Donati}, J. -F. and {Forveille}, T. and {Delfosse}, X. and {Dobler}, W. and {Petit}, P. and {Jardine}, M.~M. and {Collier Cameron}, A. and {Albert}, L. and {Manset}, N. and {Dintrans}, B. and {Chabrier}, G. and {Valenti}, J.~A.},
        title = "{The stable magnetic field of the fully convective star V374 Peg}",
      journal = {\mnras},
         year = 2008,
        month = feb,
       volume = {384},
       number = {1},
        pages = {77-86},
          doi = {10.1111/j.1365-2966.2007.12709.x},
archivePrefix = {arXiv},
       eprint = {0711.1418},
 primaryClass = {astro-ph},
       adsurl = {https://ui.adsabs.harvard.edu/abs/2008MNRAS.384...77M}
}

@ARTICLE{Nishikawa_Kusano_2002ApJ,
       author = {{Nishikawa}, N. and {Kusano}, K.},
        title = "{Effect of Density Stratification on the Thermal Convection in a Rotating Spherical Shell}",
      journal = {\apj},
         year = 2002,
        month = dec,
       volume = {581},
       number = {1},
        pages = {745-759},
          doi = {10.1086/344138},
       adsurl = {https://ui.adsabs.harvard.edu/abs/2002ApJ...581..745N}
}

@ARTICLE{Brown_et_al_ApJ_2008,
       author = {{Brown}, Benjamin P. and {Browning}, Matthew K. and {Brun}, Allan Sacha and {Miesch}, Mark S. and {Toomre}, Juri},
        title = "{Rapidly Rotating Suns and Active Nests of Convection}",
      journal = {\apj},
         year = 2008,
        month = dec,
       volume = {689},
       number = {2},
        pages = {1354-1372},
          doi = {10.1086/592397},
archivePrefix = {arXiv},
       eprint = {0808.1716},
 primaryClass = {astro-ph},
       adsurl = {https://ui.adsabs.harvard.edu/abs/2008ApJ...689.1354B}
}

@ARTICLE{Vasil_Nature_2024,
       author = {{Vasil}, Geoffrey M. and {Lecoanet}, Daniel and {Augustson}, Kyle and {Burns}, Keaton J. and {Oishi}, Jeffrey S. and {Brown}, Benjamin P. and {Brummell}, Nicholas and {Julien}, Keith},
        title = "{The solar dynamo begins near the surface}",
      journal = {\nat},
         year = 2024,
        month = may,
       volume = {629},
       number = {8013},
        pages = {769-772},
          doi = {10.1038/s41586-024-07315-1},
archivePrefix = {arXiv},
       eprint = {2404.07740},
 primaryClass = {astro-ph.SR},
       adsurl = {https://ui.adsabs.harvard.edu/abs/2024Natur.629..769V}
}

@ARTICLE{Brandenburg_et_al_ApJ_2025,
       author = {{Brandenburg}, Axel and {Larsson}, Gustav and {Del Sordo}, Fabio and {K{\"a}pyl{\"a}}, Petri J.},
        title = "{Magnetorotational Instability in a Solar Near-surface Mean-field Dynamo}",
      journal = {\apj},
         year = 2025,
        month = nov,
       volume = {993},
       number = {1},
          eid = {22},
        pages = {22},
          doi = {10.3847/1538-4357/ae03c4},
archivePrefix = {arXiv},
       eprint = {2504.16849},
 primaryClass = {astro-ph.SR},
       adsurl = {https://ui.adsabs.harvard.edu/abs/2025ApJ...993...22B}
}

@ARTICLE{Zweibel_Nature_2024,
       author = {{Zweibel}, Ellen},
        title = "{Instability could explain the Sun's curious cycle}",
      journal = {\nat},
         year = 2024,
        month = may,
       volume = {629},
       number = {8013},
        pages = {762-763},
          doi = {10.1038/d41586-024-01357-1},
       adsurl = {https://ui.adsabs.harvard.edu/abs/2024Natur.629..762Z}
}
\bibliographystyle{aasjournal}
%
\end{document}